\documentclass[onecolumn,nofootinbib,notitlepage,11pt,aps,pra,tightenlines]{revtex4-1}

\usepackage[ colorlinks = true,
             linkcolor = blue,
             urlcolor  = blue,
             citecolor = red,
             anchorcolor = green,
]{hyperref}

\usepackage[utf8]{inputenc}  
\usepackage{ifthen}
\usepackage{xparse}

\makeatletter

\DeclareUnicodeCharacter{03B1}{\ensuremath{\alpha}}
\DeclareUnicodeCharacter{03B2}{\ensuremath{\beta}}
\DeclareUnicodeCharacter{03B3}{\ensuremath{\gamma}}
\DeclareUnicodeCharacter{0393}{\ensuremath{\Gamma}}
\DeclareUnicodeCharacter{03B4}{\ensuremath{\delta}}
\DeclareUnicodeCharacter{0394}{\ensuremath{\Delta}}
\DeclareUnicodeCharacter{03B5}{\ensuremath{\varepsilon}}
\DeclareUnicodeCharacter{03B6}{\ensuremath{\zeta}}
\DeclareUnicodeCharacter{03B7}{\ensuremath{\eta}}
\DeclareUnicodeCharacter{03B8}{\ensuremath{\vartheta}}
\DeclareUnicodeCharacter{0398}{\ensuremath{\Theta}}
\DeclareUnicodeCharacter{03BA}{\ensuremath{\kappa}}
\DeclareUnicodeCharacter{03BB}{\ensuremath{\lambda}}
\DeclareUnicodeCharacter{039B}{\ensuremath{\Lambda}}
\DeclareUnicodeCharacter{00B5}{\ensuremath{\mu}}      
\DeclareUnicodeCharacter{03BC}{\ensuremath{\mu}}
\DeclareUnicodeCharacter{03BD}{\ensuremath{\nu}}
\DeclareUnicodeCharacter{03BE}{\ensuremath{\xi}}
\DeclareUnicodeCharacter{039E}{\ensuremath{\Xi}}
\DeclareUnicodeCharacter{03B9}{\ensuremath{\iota}}
\DeclareUnicodeCharacter{03C0}{\ensuremath{\pi}}
\DeclareUnicodeCharacter{03A0}{\ensuremath{\Pi}}
\DeclareUnicodeCharacter{03C1}{\ensuremath{\rho}}
\DeclareUnicodeCharacter{03C3}{\ensuremath{\sigma}}
\DeclareUnicodeCharacter{03A3}{\ensuremath{\Sigma}}
\DeclareUnicodeCharacter{03C4}{\ensuremath{\tau}}
\DeclareUnicodeCharacter{03C6}{\ensuremath{\varphi}}
\DeclareUnicodeCharacter{03A6}{\ensuremath{\Phi}}
\DeclareUnicodeCharacter{03C7}{\ensuremath{\chi}}
\DeclareUnicodeCharacter{03C8}{\ensuremath{\psi}}
\DeclareUnicodeCharacter{03A8}{\ensuremath{\Psi}}
\DeclareUnicodeCharacter{03C9}{\ensuremath{\omega}}
\DeclareUnicodeCharacter{03A9}{\ensuremath{\Omega}}
\DeclareUnicodeCharacter{03C5}{\ensuremath{\upsilon}}
\DeclareUnicodeCharacter{03A5}{\ensuremath{\Upsilon}}

\DeclareUnicodeCharacter{1FF6}{\ensuremath{\tilde{\omega}}}

\DeclareUnicodeCharacter{00AC}{\@specialsub}
\DeclareUnicodeCharacter{2042}{\@align}
\newcommand{\@align}[2][\@empty]
{\ifx\@empty#1%
  \begin{align*}#2\end{align*}%
\else
  \ifthenelse{\equal{#1}{}}%
  {\begin{align}#2\end{align}}%
  {\begin{align}#2\gdef\last@label{eq:#1}\label{eq:#1}\end{align}}
\fi}
\DeclareUnicodeCharacter{00B6}{\@eqlabel}
\newcommand*{\@eqlabel}[1][\@empty]
{ \ifx\@empty#1
    { \nonumber } 
  \else
    { \gdef\last@label{eq:#1} \label{eq:#1} }
  \fi
}
\DeclareUnicodeCharacter{00A7}{\@eqref}
\newcommand*{\@eqref}[1][\@empty]
{\ifx\@empty#1{\eqref{\last@label} }\else{\eqref{eq:#1}}\fi}

\DeclareUnicodeCharacter{221A}{\sqrt}
\DeclareUnicodeCharacter{2264}{\leq}
\DeclareUnicodeCharacter{2265}{\geq}
\DeclareUnicodeCharacter{221E}{\infty}
\DeclareUnicodeCharacter{2211}{\sum}
\DeclareUnicodeCharacter{2208}{\in}
\DeclareUnicodeCharacter{2209}{\notin}
\DeclareUnicodeCharacter{2202}{\partial}
\DeclareUnicodeCharacter{222B}{\int}
\DeclareUnicodeCharacter{2148}{\id}  
\DeclareUnicodeCharacter{2248}{\approx}  
\DeclareUnicodeCharacter{2260}{\neq}  
\DeclareUnicodeCharacter{00B1}{\pm}  
\DeclareUnicodeCharacter{2A02}{\otimes}
\DeclareUnicodeCharacter{2A01}{\oplus}
\DeclareUnicodeCharacter{00BD}{\nicefrac{1}{2}}
\DeclareUnicodeCharacter{00D7}{\times}
\DeclareUnicodeCharacter{00B7}{\cdot}
\DeclareUnicodeCharacter{2032}{\ensuremath{'}}
\DeclareUnicodeCharacter{2020}{^{\dagger}}
\DeclareUnicodeCharacter{1D40}{^{\textnormal{\tiny{T}}}}
\DeclareUnicodeCharacter{00B2}{^{2}}
\DeclareUnicodeCharacter{00B3}{^{3}}
\DeclareUnicodeCharacter{203E}{\@bar}
\newcommand{\@bar}{\bar}
\DeclareUnicodeCharacter{2040}{\@hat}
\newcommand{\@hat}{\hat}
\DeclareUnicodeCharacter{2329}{\langle}
\DeclareUnicodeCharacter{232A}{\rangle}
\DeclareUnicodeCharacter{27E8}{\langle}
\DeclareUnicodeCharacter{27E9}{\rangle}
\DeclareUnicodeCharacter{2016}{\|}
\DeclareUnicodeCharacter{2192}{\to}
\DeclareUnicodeCharacter{21D2}{\implies}
\DeclareUnicodeCharacter{21A6}{\mapsto}
\DeclareUnicodeCharacter{2113}{\ell}
\DeclareUnicodeCharacter{210B}{\h}
\DeclareUnicodeCharacter{2115}{\bN}
\DeclareUnicodeCharacter{211D}{\bR}
\DeclareUnicodeCharacter{2102}{\bC}
\DeclareUnicodeCharacter{2119}{\Pr}
\DeclareUnicodeCharacter{2130}{\sE}
\DeclareUnicodeCharacter{2133}{\cM}
\DeclareUnicodeCharacter{261B}{\item} 

\DeclareUnicodeCharacter{00E1}{\'a}

\DeclareUnicodeCharacter{263C}{\@multiindex} 
\NewDocumentCommand{\@multiindex}{s o m}
{\IfBooleanTF#1%
{\index{#3}}%
{\IfNoValueTF{#2}%
{\emph{#3}\index{#3}}%
{\emph{#2}\index{#3}}}}

\makeatother

\usepackage[sans]{dsfont} 
\usepackage{bbm}
\usepackage[mathscr]{euscript}
\usepackage{amsmath}
\usepackage{amssymb}
\usepackage{amsthm}
\usepackage{graphicx}
\usepackage{units}
\usepackage{ifthen}
\usepackage{xparse}

\makeatletter

\theoremstyle{plain}
\newtheorem{lemma}{Lemma}
\newtheorem{proposition}[lemma]{Proposition}

\newtheorem{corollary}[lemma]{Corollary}

\newtheorem{definition}{Definition}

\theoremstyle{definition}

\newcommand*{\@specialsub}[1]{_{\textnormal{\tiny $#1$}}}
\newcommand*{\ii}[2]{{#1}\@specialsub{#2}}

\newcommand*{\mod@def}[1]{
 \def\mod@l{} \def\mod@r{}
 \ifthenelse{\equal{#1}{}} {} 
 {\ifthenelse{\equal{#1}{n}} {} 
 {\ifthenelse{\equal{#1}{?}} {\def\mod@l{\left} \def\mod@r{\right}} 
 {\ifthenelse{\equal{#1}{*}} {\def\mod@l{\left} \def\mod@r{\right}} 
 {\ifthenelse{\equal{#1}{b}} {\def\mod@l{\big} \def\mod@r{\big}} 
 {\ifthenelse{\equal{#1}{B}} {\def\mod@l{\Big} \def\mod@r{\Big}} 
 {\ifthenelse{\equal{#1}{g}} {\def\mod@l{\bigg} \def\mod@r{\bigg}} 
 {\ifthenelse{\equal{#1}{G}} {\def\mod@l{\Bigg} \def\mod@r{\Bigg}} 
   {\typeout{LaTeX Warning: Unknown size modifier: '#1' on line \the\inputlineno.}}
}}}}}}}}
\newcommand*{\mod@defnostar}[1]{
 \def\mod@a{}
 \ifthenelse{\equal{#1}{}} {} 
 {\ifthenelse{\equal{#1}{n}} {} 
 {\ifthenelse{\equal{#1}{b}} {\def\mod@a{\big}} 
 {\ifthenelse{\equal{#1}{B}} {\def\mod@a{\Big}} 
 {\ifthenelse{\equal{#1}{g}} {\def\mod@a{\bigg}} 
 {\ifthenelse{\equal{#1}{G}} {\def\mod@a{\Bigg}} 
   {\typeout{LaTeX Warning: Unknown size modifier: '#1' on line \the\inputlineno.}}
}}}}}}

\NewDocumentCommand \abs { D<>{n} m }
{ \mod@def{#1} 
   \mod@l| {#2} \mod@r|
}
\NewDocumentCommand \norm { s D<>{n} m O{} }
{ \mod@def{#2} 
  \IfBooleanTF#1
    { \mod@l|\!\mod@l|\!\mod@l| {#3} \mod@r|\!\mod@r|\!\mod@r|_{#4} }
    { \mod@l|\!\mod@l| {#3} \mod@r|\!\mod@r|_{#4} }
}

\DeclareMathOperator{\trace}{tr}
\NewDocumentCommand \tr { s D<>{n} O{} m } 
{ \IfBooleanTF#1
     { \mod@def{?} } 
     { \mod@def{#2} } 
  \ifthenelse{\equal{#3}{}}
    { \trace \mod@l( {#4} \mod@r) }
    { \ii{\trace}{#3} \mod@l( {#4} \mod@r) }
}

\DeclareMathOperator{\@kern}{kern}
\RenewDocumentCommand \ker { D<>{n} m }
  { \def\cmd@mathcmd{\@kern\,} \@mathcmd{#1}{#2} }

\DeclareMathOperator{\@imag}{image}
\NewDocumentCommand \imag { D<>{n} m }
  { \def\cmd@mathcmd{\@imag\,} \@mathcmd{#1}{#2} }

\DeclareMathOperator{\@supp}{supp}
\NewDocumentCommand \supp { D<>{n} m }
  { \def\cmd@mathcmd{\@supp\,} \@mathcmd{#1}{#2} }

\DeclareMathOperator{\@rank}{rank}
\NewDocumentCommand \rank { D<>{n} m }
  { \def\cmd@mathcmd{\@rank\,} \@mathcmd{#1}{#2} }
  
\DeclareMathOperator{\@dim}{dim}
\RenewDocumentCommand \dim { D<>{n} m }
  { \def\cmd@mathcmd{\@dim\,} \@mathcmd{#1}{#2} }

\DeclareMathOperator{\@vecspan}{span}
\NewDocumentCommand \vecspan { D<>{n} m }
  { \def\cmd@mathcmd{\@vecspan\,} \@mathcmd{#1}{#2} }

\newcommand*{\@mathcmd}[2]
{ \mod@def{#1}
    { \cmd@mathcmd \mod@l\{ { #2 } \mod@r\} }
}

\NewDocumentCommand \mat { s m +m }
{ \IfBooleanTF#1
  { \left[\begin{array}{#2} #3 \end{array}\right] }
  { \left(\begin{array}{#2} #3 \end{array}\right) }
}

\NewDocumentCommand \ip { D<>{n} >{\SplitArgument{1}{,}}m }
  { \mod@def{#1} \@ip#2 }
\newcommand*{\@ip}[2]{\mod@l\langle {#1}, {#2} \mod@r\rangle}
\NewDocumentCommand \bra { D<>{n} m O{} }
{ \mod@def{#1}
  \ifthenelse{\equal{#3}{}}
  { \mod@l\langle {#2} \mod@r| }
  { \ii{\mod@l\langle {#2} \mod@r|}{#3} }
}
\NewDocumentCommand \ket { D<>{n} m O{} }
{ \mod@def{#1}
  \ifthenelse{\equal{#3}{}}
  { \mod@l| {#2} \mod@r\rangle }
  { \ii{\mod@l| {#2} \mod@r\rangle}{\!#3} }
}
\NewDocumentCommand \proj { D<>{n} m O{} }
{ \mod@def{#1}
  \ifthenelse{\equal{#3}{}}
  { \mod@l| {#2} \mod@r\rangle\!\mod@l\langle {#2} \mod@r| }
  { \mod@l| {#2} \mod@r\rangle\!\mod@l\langle {#2} \mod@r|\@specialsub{#3} }
}
\NewDocumentCommand \braket { D<>{n} >{\SplitArgument{2}{|}} m O{} }
{ \mod@defnostar{#1} 
  \ifthenelse{\equal{#3}{}} {\@braket#2} {\ii{\@braket#2}{#3}}
}
\newcommand*{\@braket}[3]
{ 
  \IfValueTF{#3}     
  { \mod@a\langle #1 \mod@a| {#2} \mod@a| {#3} \mod@a\rangle }
  { \mod@a\langle #1 \mod@a| {#2} \mod@a\rangle }
}

\NewDocumentCommand \tn { t- s m }
{ \IfBooleanTF#1
  { \textnormal{#3} }
  { \IfBooleanTF#2
    { \quad \textnormal{#3} \quad }
    { \ \textnormal{#3}\ }
  }
}

\NewDocumentCommand \h { O{} }
  { \ifthenelse{\equal{#1}{}} {\sH} {\ii{\sH}{#1}} }
\NewDocumentCommand \id { s O{} }
{ \IfBooleanTF#1
  { \ifthenelse{\equal{#2}{}} {\mathbbmtt{i}} {\ii{\mathbbmtt{i}}{#2}} }
  { \ifthenelse{\equal{#2}{}} {\mathbbmtt{1}} {\ii{\mathbbmtt{1}}{#2}} }
}
\NewDocumentCommand \opid { O{} }
  { \ifthenelse{\equal{#1}{}} {\cI} {\ii{\cI}{#1}} }

\NewDocumentCommand \ball { s D<>{n} O{\eps} m }
{ \mod@def{#2}
  \IfBooleanTF#1
  { \cB_{\textnormal{p}}^{#3}\mod@l( #4 \mod@r) } 
  { \cB^{#3} \mod@l( #4 \mod@r) }
}

\NewDocumentCommand{\hh} { s D<>{n} m O{} >{\SplitArgument{1}{|}}m O{} }
  { \@hh{#1}{#2}{#3}{#4}#5{#6} }
\NewDocumentCommand{\hmin} { t+ s D<>{n} O{} >{\SplitArgument{1}{|}}m O{} }
{ \IfBooleanTF#1 
  { \@hhalt{#2}{#3}{\min}{#4}#5{#6} }
  { \@hh{#2}{#3}{\min}{#4}#5{#6} } 
}
\NewDocumentCommand{\hminsymb} { t+ s }
{ \IfBooleanTF#1 {\widehat{H}}{H}
  ^{\IfBooleanTF#2 {\eps}{}}
  _{\min}
}
\NewDocumentCommand{\hmax} { t+ s D<>{n} O{} >{\SplitArgument{1}{|}}m O{} }
{ \IfBooleanTF#1 
  { \@hhalt{#2}{#3}{\max}{#4}#5{#6} }
  { \@hh{#2}{#3}{\max}{#4}#5{#6} }
}
\NewDocumentCommand{\hvn} { D<>{n} >{\SplitArgument{1}{|}}m O{} }
  { \@hh{\BooleanFalse}{#1}{}{}#2{#3} }
  
\newcommand*{\@hh}[7]
{ \mod@defnostar{#2}
  \IfBooleanTF{#1} { \def\arg@eps{\eps} } { \def\arg@eps{#4} }
  \IfValueTF{#6}
    { H_{#3}^{\arg@eps} \mod@a( #5 \mod@a| 
      #6 \mod@a)_{#7} }
    { H_{#3}^{\arg@eps} \mod@a( #5 \mod@a)_{#7} }
}
\newcommand*{\@hhalt}[7]
{ \mod@defnostar{#2}
  \IfBooleanTF{#1} { \def\arg@eps{\eps} } { \def\arg@eps{#4} }
  \IfValueTF{#6}
    { \widehat{H}_{#3}^{\arg@eps} \mod@a( #5 \mod@a| 
      #6 \mod@a)_{#7} }
    { \widehat{H}_{#3}^{\arg@eps} \mod@a( #5 \mod@a)_{#7} }
}

\newcommand*{\eps}{\varepsilon}

\newcommand{\cA}{\mathcal{A}}
\newcommand{\cB}{\mathcal{B}}

\newcommand{\cE}{\mathcal{E}}
\newcommand{\cF}{\mathcal{F}}
\newcommand{\cG}{\mathcal{G}}
\newcommand{\cH}{\mathcal{H}}
\newcommand{\cI}{\mathcal{I}}
\newcommand{\cJ}{\mathcal{J}}
\newcommand{\cK}{\mathcal{K}}

\newcommand{\cM}{\mathcal{M}}
\newcommand{\cN}{\mathcal{N}}

\newcommand{\cP}{\mathcal{P}}

\newcommand{\cS}{\mathcal{S}}
\newcommand{\cT}{\mathcal{T}}

\newcommand{\sE}{\mathscr{E}}

\newcommand{\sH}{\mathscr{H}}

\newcommand{\bN}{\mathbb{N}}
\newcommand{\bR}{\mathbb{R}}
\newcommand{\bC}{\mathbb{C}}

\makeatother

\let\originalleft\left
\let\originalright\right
\renewcommand{\left}{\mathopen{}\mathclose\bgroup\originalleft}
\renewcommand{\right}{\aftergroup\egroup\originalright}

\def\H{{\mathcal H}}
\def\BB{{\mathcal B}}
\def\EE{{\mathcal E}}
\def\DD{{\mathcal D}}
\def\P{{\mathcal P}}
\def\M{{\mathcal M}}

\def\N{{\mathcal N}}
\def\F{{\mathcal F}}
\def\C{{\mathbb C}}
\def\E{{\mathbb E}}
\def\S{{\mathcal S}}
\def\idty{{\leavevmode\rm 1\mkern -5.4mu I}}
\newcommand\Scp[2]{\ensuremath{\, \langle #1 \,\vert #2 \,\rangle}}
\newcommand*{\vphi}{\varphi}
\def\tr{{\rm Tr}}

\newcommand\Dmax[2]{\ensuremath{\mathrm{D}_{\max} \left(#1 \left|\right| #2 \right)}}
\newcommand\Dmin[2]{\ensuremath{\mathrm{D}_{\min} \left(#1 \left|\right| #2 \right)}}

\newcommand\Hmin[2]{\ensuremath{\mathrm{H}_{\min} \left(#1 \vert #2 \right)}}
\def\Ss{{\S_{\leq}}}
\newcommand\Hmax[2]{\ensuremath{\mathrm{H}_{\max} \left(#1 \vert #2 \right)}}
\def\pemb{{\curvearrowright}}
\newcommand\SHmin[2]{\ensuremath{\mathrm{H}^{\epsilon}_{\min} \left(#1 \vert #2 \right)}}
\newcommand\SHmax[2]{\ensuremath{\mathrm{H}^{\epsilon}_{\max} \left(#1 \vert #2 \right)}}
\newcommand\Ball[2]{\ensuremath{\cB^{\epsilon}_{#1}(#2)}}
\def\ketbra #1#2{\vert #1\rangle \langle #2\vert}
\def\norm #1{\Vert #1\Vert}

\usepackage{color}
\definecolor{myred}{rgb}{1,0,0}

\definecolor{myblue}{rgb}{0,0,0.8}

\definecolor{myyellow}{rgb}{0.9,0.8,0}

\definecolor{mygreen}{rgb}{0,0.6,0}

\definecolor{myorange}{rgb}{0.6,0.6,0}

\definecolor{mycerul}{rgb}{0,0.6,1}

\begin{document}

\title{The Smooth Entropy Formalism for von Neumann Algebras}

\author{Mario Berta}
\email[]{berta@caltech.edu}
\affiliation{Institute for Quantum Information and Matter, California Institute of Technology, USA}

\author{Fabian Furrer}
\email[]{furrer@eve.phys.s.u-tokyo.ac.jp}
\affiliation{Department of Physics, Graduate School of Science, University of Tokyo, Japan}
\affiliation{Institute for Theoretical Physics, Leibniz University Hanover, Germany}

\author{Volkher B.~Scholz}
\email[]{scholz@phys.ethz.ch}
\affiliation{Institute for Theoretical Physics, ETH Zurich, Switzerland}

\begin{abstract}
We discuss information-theoretic concepts on infinite-dimensional quantum systems. In particular, we lift the smooth entropy formalism as introduced by Renner and collaborators for finite-dimensional systems to von Neumann algebras. For the smooth conditional min- and max-entropy we recover similar characterizing properties and information-theoretic operational interpretations as in the finite-dimensional case. We generalize the entropic uncertainty relation with quantum side information of Tomamichel and Renner and discuss applications to quantum cryptography. In particular, we prove the possibility to perform privacy amplification and classical data compression with quantum side information modeled by a von Neumann algebra.
\end{abstract}

\maketitle


\section{Introduction}\label{sec:intro}

During the last decades many concepts and techniques have been developed to study quantum information-theoretic tasks using physical systems described by finite-dimensional Hilbert spaces (see, e.g, the books~\cite{wildebook13,NieChu00Book}). One conceptually interesting building block is the smooth entropy formalism as introduced by Renner and collaborators~\cite{renner05,mybook}. In this work, we extend its scope to more general physical systems modeled by von Neumann algebras. The general aim is the development of a mathematical framework suited to describe quantum informational tasks with resources like bosonic or fermionic quantum fields (see, e.g., the books~\cite{Haag92,ruelle1999statistical} for further discussions on the algebraic formulation of quantum fields).

A fundamental concept in classical and quantum information theory are entropy measures. They can be defined via an axiomatic approach~\cite{renyi60}, or operationally, in the sense that they quantitatively characterize fundamental tasks in information theory~\cite{Shannon48}. If the resources are independent and identically distributed, the relevant measures in the asymptotic limit turn out to be the von Neumann entropy~\cite{vonneumann32} and Umegaki's relative entropy~\cite{umegaki62}. The definition of these entropies in the setup of von Neumann algebras and the investigation of their properties are closely connected to developments in the algebraic formulation of quantum theory. Early contributors, among others, are Araki, Benatti, Connes, Fannes, Narnhofer, Petz, Thirring, and Uhlmann (see, e.g, the book~\cite{Petz_Book} and references therein).

In order to analyze resources of general form, Renner and collaborators developed the smooth entropy formalism (see, e.g., \cite{renner05,mybook} and references therein). The fundamental entropic quantities are the smooth conditional min- and max-entropy, which characterize the optimal performance of basic information-theoretic tasks for arbitrary resources. For independent and identically distributed resources, the conditional von Neumann entropy is recovered in the asymptotic limit of infinitely many repetitions~\cite[Theorem 1]{Tomamichel08}.

In this paper, we extend the smooth entropy formalism to the algebraic approach of quantum mechanics. This enables to study information-theoretic problems with infinite-dimensional quantum systems, like for instance quantum fields of bosons and fermions
or other continuous variable systems (see, e.g., the review article~\cite{weedbrook11} and references therein). In the special case that the von Neumann algebra is equal to the algebra of all linear bounded operators on some separable Hilbert space, the smooth entropy formalism has been studied in~\cite{Furrer10}. It was shown that many results from the finite-dimensional case carry over via an inductive limit taken over all finite-dimensional subspaces. However, the assumption of a full algebra is often too restrictive. For example, the von Neumann algebra of a field of free bosons at finite temperature is not of this type~\cite{ArakiWoods}.\\

Let us briefly summarize how the paper is organized. We start in Section~\ref{sec:notation} with a brief introduction to von Neumann algebras, followed by a discussion of the relevant quantum information-theoretic concepts (Section~\ref{sec:setup}). We then proceed in Section~\ref{sec:cond_minmax} with the definition of the conditional min- and max-entropy. In Sections~\ref{sec:purified} and~\ref{sec:smoothcond}, we define and discuss the smooth conditional min- and max-entropy. This is followed by a discussion of their properties (Section~\ref{sec:properties_smooth}), as well as an extension to min- and max-relative entropy (Section~\ref{sec:relative_minmax}). Finally, we discuss applications in quantum information theory (Section~\ref{sec:qit}). This includes the operational meaning of the conditional min- and max-entropy (Section~\ref{sec:cond}), the special case of classical quantum systems (Section~\ref{sec:classical_quantum}), uncertainty relations for the smooth conditional min- and max-entropy (Section~\ref{sec:uncertainty}), as well as applications in quantum cryptography (Section~\ref{sec:qkd_applications}). We end with a summary of our results and a presentation of some perspectives concerning applications (Section~\ref{sec:outlook}).


\section{Preliminaries}\label{sec:prel}

Here we recall some basic concepts and mathematical tools needed to describe quantum information theory in the framework of von Neumann algebras. For an introduction to the theory of von Neumann algebras we refer the reader to the books~\cite{Bratteli1,Takesaki1}.


\subsection{Mathematical Background}\label{sec:notation}

\textbf{$\mathbf{C^*}$-algebras.} A $*$-algebra is an algebra $\cA$, which is also a vector space over $\mathbb{C}$, together with an operation $*$ called involution satisfying $A^{**}=A$, $(AB)^{*}=B^{*}A^{*}$ and $(\alpha A+\beta B)^{*}=\bar{\alpha}A^{*}+\bar{\beta}B^{*}$ for all $A,B\in\cA$ and $\alpha,\beta\in\mathbb C$. If a $*$-algebra is equipped with a sub-multiplicative norm for which the involution is isometric and the algebra complete, it is called a Banach $*$-algebra.

\begin{definition}\label{def:cstar}
A $C^{*}$-algebra is a Banach $*$-algebra $\cA$ with the property
\begin{align}
\|A^{*}A\|=\|A\|^{2}\,,
\end{align}
for all $A\in\cA$.
\end{definition}

Note that the set of all linear, bounded operators on a Hilbert space $\H$, denoted by $\cB(\H)$, is a $C^*$-algebra with the usual operator norm (induced by the norm on $\cH$), and the adjoint operation. Furthermore, each norm closed $*$-subalgebra of $\cB(\H)$ is a $C^*$-algebra. A representation of a $C^*$-algebra $\cA$ is a $*$-homomorphism $\pi:\cA\rightarrow\cB(\H)$ on a Hilbert space $\H$. A $*$-homomorphism is a linear map compatible with the $*$-algebraic structure, that is, $\pi(AB)=\pi(A)\pi(B)$ and $\pi(A^*)=\pi(A)^*$. We call a representation $\pi$ faithful if it is an isometry, which is equivalent to say that it is a $*$-isomorphism from $\cA$ to $\pi(\cA)$. A basic theorem in the theory of $C^*$-algebras says that each $\cA$ is isomorphic to a norm closed $*$-subalgebra of a $\cB(\H)$ with suitable $\cH$~\cite[Theorem 2.1.10]{Bratteli1}. Hence, each $C^*$-algebra can be seen as a norm closed $*$-subalgebra of a $\cB(\H)$.

An element $b \in \cA$ is called positive if  $b=a^*a$ for $a\in\cA$, and the set of all positive elements is denoted by $\cA_+$.
A linear functional $\omega$ in the dual space $\cA^*$ of $\cA$ is called positive if $\omega(a)\geq 0$ for all $a\in\cA_+$. The set of all positive functionals $\cA^*_+$ defines a positive cone in $\cA^*$ with the usual ordering $\omega_{1}\geq\omega_{2}$ if $(\omega_{1}-\omega_{2})\in\cA^*_+$, and we say that $\omega_{1}$ majorizes $\omega_{2}$. The norm on the dual space of $\cA$ is defined as
\begin{align}\label{def:normDualspace}
\Vert \omega\Vert:=\sup_{x\in\cA , \Vert x\Vert\leq 1} \vert \omega(x)\vert\,.
\end{align}
A positive functional $\omega\in\cA^*$ with $\Vert \omega\Vert = 1$ is called a state. A state $\omega$ is called pure if the only positive linear functionals which are majorized by $\omega$ are given by $\lambda\cdot\omega$ for $0\leq\lambda\leq1$. If $\cA=\cB(\H)$ we have that the pure states are exactly the functionals $\omega_{\xi}(x)=\Scp {\xi}{x\xi}$, where $\ket \xi \in \H$.\\

\textbf{Von Neumann algebras.} We consider a subset of linear bounded operators $\cT\subset\cB(\H)$ on a Hilbert space $\cH$. The commutant $\cT'$ of $\cT$ is defined as $\cT'=\{a\in\cB(\H):[a,x]=0,\forall \, x\in\cT\}$, where $[a,x]:=ax-xa$.  

\begin{definition}\label{def:neumann}
Let $\H$ be a Hilbert space. A von Neumann algebra $\M$ acting on $\H$ is a $*$-subalgebra $\M\subset\cB(\H)$ which satisfies $\M''=\M$.
\end{definition}

Beside the above definition, there exist other ways to characterize a von Neumann algebra. One rises in the bicommutant theorem \cite[Lemma 2.4.11]{Bratteli1}: a $*$-subalgebra $\M\subset\cB(\H)$ containing the identity is $\sigma$-weakly closed if and only if $\M''=\M$. (The $\sigma$-weak topology on $\cB(\H)$ is the locally convex topology induced by the semi-norms $A \mapsto \vert \tr (\tau A)\vert$ for trace-class operators $\tau\in\cB(\H)$, see~\cite[Chapter 2.4.1]{Bratteli1}.) From this we can conclude that a von Neumann algebra $\M$ is also norm closed and therefore a $C^*$-algebra. We note that a norm closed subalgebra is not necessarily $\sigma$-weakly closed. Thus, a $C^*$-algebra on $\H$ is not always a von Neumann algebra. The definition of a von Neumann algebra can even be stated in the category of $C^*$-algebras: a von Neumann algebra $\cM$ is a $C^*$-algebra with the property that it is the dual space of a Banach space. Due to historical reasons this is also called a $W^*$-algebra.

In the following $\cM$ denotes a von Neumann algebra. A representation $\pi$ of a von Neumann algebra $\M$ is a $*$-representation on a Hilbert space $\H$ that is $\sigma$-weakly continuous. Thus, the image $\pi(\M)$ is again a von Neumann algebra. We say that two von Neumann algebras are isomorphic if there exists a faithful representation mapping one into the other.

Given two commuting von Neumann algebras $\M$ and $\hat\M$ acting on the same Hilbert space $\H$, we define the von Neumann algebra generated by $\M$ and $\hat\M$ as $\M\vee\hat\M=(\M\cup\hat\M)''$, where $\M\cup\hat\M=\rm{span}\{xy\; ;\; x\in\M \, , y\in\hat\M\}$. According to the bicommutant theorem \cite[Lemma 2.4.11]{Bratteli1}, $\M\vee\hat\M$ is just the $\sigma$-weak closure of $\M\cup\hat\M$.\\

\textbf{Functionals on von Neumann algebras.} A linear functional $\omega:\cM\rightarrow\mathbb{C}$ is called normal if for any monotone increasing net of operators \(x_\alpha\in\cM\) with least upper bound \(x\), \(\omega(x_\alpha)\) converges to \(\omega(x)\). Equivalently, it is $\sigma$-weakly continuous~\cite[Chapter 7, Theorem 7.1.12]{kadison1986fundamentals}. We denote the set of linear, normal functionals on $\cM$ by $\cN(\cM)$.

We equip $\cN(\M)$ with the usual norm as given in (\ref{def:normDualspace}). Then the set $\cN(\M)$ is a Banach space and moreover it is the predual of $\M$, which means that its dual space is $\M$. The cone of positive elements in  $\cN(\cM)$ is denoted by $\cN^{+}(\cM)$. We have that $\|\omega\| =\omega(\idty)$ for all $\omega\in\cN^{+}(\cM)$, where $\idty$ denotes the identity element in $\cM$. We call functionals $\omega\in\cN^+(\cM)$ with $\|\omega\|\leq1$ subnormalized states and denote the set of all subnormalized states by $\cS_{\leq}(\cM)$. Moreover, we say that $\omega\in\cS_{\leq}(\cM)$ is a normalized state if $\|\omega\|=1$, and set
\begin{align}
\cS(\cM):=\{\omega\in\cN^+(\cM):\|\omega\|=1\}\,.
\end{align}
For $\M\subset\cB(\H)$, we have that for any $\omega\in \cN^+(\cM)$ exists a positive trace-class operator $\rho$ on $\cH$, such that
\begin{align}\label{eq:densityMatrix}
\omega_{\rho}(x)=\tr(\rho x) = \omega(x)\quad\forall x\in\M\,.
\end{align}
If $\omega$ is normalized, such an operator $\rho$ is called a density operator. A particular example is a vector state $\omega_\xi(x) = \Scp{\xi}{x \xi}$, given by some unit vector $\ket{\xi} \in \H$. The Gelfand-Naimark-Segal (GNS) construction~\cite{gelfand43,segal47} asserts that for every state $\omega$ there exists a Hilbert space $\H_\omega$, together with a unit vector $\ket{\xi_\omega} \in \H_\omega$ and a representation $\pi_\omega:\cM \to \BB(\H)$ such that $\omega = \omega_{\xi_\omega} \circ \pi_\omega$, i.e.,
\begin{align}
\omega(x)=\langle\xi_\omega|\pi_\omega(x) \xi_\omega\rangle\quad\forall x\in\cM\,.
\end{align}
Moreover, the vector $\ket{\xi_\omega}$ is cyclic, that is, $\H_\omega$ is the closure of $\{\pi_\omega(x)\ket{\xi_\omega} : x\in\M\}$.\\

\textbf{Weights on von Neumann algebras.} In addition to normal states, we also consider weights. A weight $\vphi$ on a von Neumann algebra $\cM$ is a map from the positive elements in $\cM$ into the positive reals, being possibly infinite, satisfying
\begin{align}
\vphi(x+y) = \vphi(x) + \vphi(y)\quad\mathrm{and}\quad\vphi(\lambda x) = \lambda \vphi(x),\quad\mathrm{for}\quad\forall x,y \in \cM_+,\;\lambda\geq0\,.
\end{align}
A weight is called semi-finite if the set $\{x \in \M_+ \,:\,\vphi(x)<\infty\}$  is $\sigma$-weakly dense in \(\M\)~\cite[Chapter VII, Definition 1.1]{Takesaki1}. It is called faithful if \(\vphi(x)\neq 0\) for any non-zero element \(x \in \M_+\). Moreover, a weight $\vphi$ is called normal if, similar to the case of linear functionals, $\vphi(x_\alpha)$ converges to $\vphi(x)$ for any monotone increasing net of operators $x_\alpha\in\cM$ with least upper bound $x$. A prime example of a semi-finite normal weight is the trace on $\BB(\H)$, with $\H$ being an infinite-dimensional Hilbert space. Here normality of a weight is defined similar as for functionals.


\subsection{Algebraic Quantum Theory}\label{sec:setup}

\textbf{Systems.} We associate to every physical system a von Neumann algebra, which is generated by the physical observables. According to Davies~\cite{Davies}, we use the most general notion of an observable and define it as a positive operator valued measure (POVM), which consists of a measurable space $(X,\Sigma)$ with $\sigma$-algebra $\Sigma$ defining the values of the possible measurement outcomes together with a $\sigma$-additive function $E: \Sigma \rightarrow \M_+$ such that $E(X)=\idty$. Henceforth, we consider only observables with a discrete outcome range $X$ described by a collection of positive operators $ \{E_x\}_{x\in X}$ in $\M$ satisfying $\sum_{x}E_x=\idty$. A measurement is called projective or of von Neumann type if the operators $E_x$ are projections. The state of a physical system is represented by a functional $\omega\in\cS(\cM)$. The probability distribution generated by a measurement described by the observable $\{E_x\}_x$ is computed via $p_x=\omega(E_x)$.\\

\textbf{Dynamics.} The possible evolution of a quantum system is described by normal completely positive unital maps $\cE:\M_B\rightarrow\M_A$. These are called quantum channels. This corresponds to a description in the Heisenberg picture (see~\cite{Paulsen} for proper definitions). The corresponding pre-dual map $\cE_*$ in the Schr\"odinger picture is defined via the relation $\cE_*(\omega)(a)=\omega(\cE(a))$ for all $a\in\M_B$ and is also completely positive. It maps $\cS(\cM_A)$ into $\cS(\cM_B)$. Note that if we consider all linear bounded operators on a Hilbert space $\cH$, a state $\omega$ is usually associated with a density matrix $\rho$ via $\omega(\cdot)=\tr\left[\cdot\rho\right]$. In this case the unitality of $\cE$ translates to $\tr\left[\cE_*(\rho)\right]=\tr\left[\rho\right]$ and is referred to as trace preserving. However, a von Neumann algebra does not always admit a trace, and this property translates to norm conservation on $\cN^+(\cM)$.\\

\textbf{Multipartite systems.} A multipartite system is a composite of different physical subsystems $A,B,\ldots,Z$ associated with mutually commuting von Neumann algebras $\M_A,\M_B,\dots,\M_Z$ acting on the same Hilbert space $\H$. (If they act on different Hilbert spaces, we just consider their action on the tensor product of the Hilbert spaces.) The corresponding von Neumann algebra of the multipartite system is given by
\begin{align}
\M_{AB\ldots Z}:=\M_A\vee\M_B\vee\ldots\vee \M_Z\,,
\end{align}
where $\M_A\vee\cM_B$ denotes the von Neumann algebra generated by $\cM_A$ and $\cM_B$. The considered subsystems are always labeled by subscripts. For example, a state on $\M_{ABC}$ is denoted by $\omega_{ABC}$ while $\omega_{AB}$ is the restriction of $\omega_{ABC}$ onto $\M_{AB}$. We remark that this characterization handles both bosonic and fermionic theories, since the von Neumann algebras correspond to observable quantities, which always commute if space-like separated (see~\cite[Chapter III.1]{Haag92} for a discussion).\\

\textbf{Purifications.} An important concept in quantum information theory is purification, which is essentially the completion of a system by adding a complementary system. The idea of purification is to choose an extension $\omega$ of a state $\omega_A$ such that $\omega$ is a pure state. The name is justified by the property that no further extension of the system shows any correlation with the purification $\omega$~\cite[Section IV, Lemma 4.11]{Takesaki1}: if $\tilde\omega\in\S(\tilde\M)$ with $\M\subset\tilde\M$ and $\tilde\omega$ restricted to $\M$ is a pure state $\omega$ on $\M$, then it follows that $\tilde\omega(xy)=\tilde\omega(x)\tilde\omega(y)$ for all $x\in\M$ and $y\in\M'\cap\tilde\M$, where $\M'$ denotes the commutant of $\cM$. 

\begin{definition}\label{def:purification}
Let $\omega\in\cS_{\leq}(\cM)$. A purification of $\omega$ is a triple $(\pi,\H, \ket \xi)$, where $\pi$ is a representation of $\cM$ on a Hilbert space $\H$ and $\xi\in\H$ such that $\omega(x)=\Scp{\xi}{\pi(x)\xi}$ for all $x\in\cM$. We call $\pi(\M)$ the relevant and $\pi(\cM)'$ the complementary system of the purification $(\pi,\H, \ket \xi)$.
\end{definition}

The GNS construction as reviewed in Section~\ref{sec:notation} can be rephrased as every state admits a purification. We say for short that $\omega_{A'B}$ is a purification of $\omega_A\in\cS_{\leq}(\cM_A)$ if there exists a purification $(\pi,\H, \ket \xi)$ of $\omega_A$ such that $\M_{A'}=\pi(\M_A)$, $\M_B= \pi(\cM_A)'$ and $\omega_{A'B}(x)=\Scp{\xi}{x\xi}$ for all $x\in\M_{A'B}$. Note that we use a less restrictive notion of purification compared to the one of Woronowicz~\cite{Woronowicz}, which only applies to factor states. This has the consequence that the state $\omega_{A'B}$ is in general not a pure state for $\cM_{A'B}$. For any von Neumann algebra $\M$, there exists a representation $\pi$ on a Hilbert space $\H$, such that every state on $\M$ has a purification in $\H$. We call such a representation a standard form of $\M$~\cite[Chapter IX.1]{Takesaki1}.\\

\textbf{Full algebras.} Special systems of interest are full algebras of all linear bounded operators on a separable Hilbert space: $\cM_A=\cB(\cH_A)$. This von Neumann algebra possesses a tracial weight \(\tau_A\) (a weight satisfying \(\tau(x^*x) = \tau(x x^*)\;\forall x\in\cM\)), which is unique if we require that it takes the value one on minimal projections. We denote this weight by \(\tau_A\), which can be identified with the usual trace on \(\cH_A\).

Moreover, we are interested in multipartite systems where only the first system is a full algebra $\cM_A=\cB(\cH_A)$. If \(\cM_B\) is another von Neumann algebra, we construct the von Neumann tensor product of \(\cM_A\) with \(\cM_B\), denoted by \(\cM_A \otimes \cM_B\). For \(\cM_B \subset \cB(\cH_B)\), the tensor product \(\cM_A \otimes \cM_B\) is the von Neumann algebra generated by the *-subalgebra \(\cM_A \otimes_{\mathrm{alg}} \cM_B \subset \cB(\cH_A \otimes \cH_B)\) with \(\otimes_{\mathrm{alg}}\) the algebraic tensor product. We briefly recall a few properties for \(\cM_A \otimes \cM_B\), a detailed discussion can be found in~\cite[Chapter IV.5]{Takesaki1}. If \(\cM_A^\prime\) and \(\M_B^\prime\) are the commutants, then we have \((\cM_A\otimes\M_B)^\prime = \M_A^\prime \otimes \M_B^\prime\). For \(\sigma \in \cS(\M_B)\) there exists a normal conditional expectation \(\hat{\sigma}:\cM_A\otimes\cM_B \to \cM_A\) such that for any \(\chi \in \cS(\M_A)\) we have \(\chi(\hat{\sigma}(a \otimes b)) = \chi(a)\sigma(b)\)~\cite[Chapter IV.5 \& Chapter IX, Theorem 4.2] {Takesaki1}. This ensures the existence of the product state \(\chi \otimes \sigma\). Likewise, we denote by \(\vphi \otimes \sigma\) the normal semi-finite weight given by \(\vphi \circ \hat{\sigma}\).

Finally, if $\M_A$ and $\M_B$ are two commuting subalgebras in some \(\cB(\cH)\) and \(\M_A\) is a full algebra (on some Hilbert space), then we can decompose the Hilbert space \(\cH = \cH_1 \otimes \cH_2\), with \(\cH_1 \simeq \cH_A\) and \(\cB(\cH_1 \otimes \cH_2) \supset \M_A \vee \M_B \simeq \cB(\cH_1) \otimes \M_B\). That is, for subsystems described by full algebras, commuting and tensor product representations agree.\\

\textbf{Finite-dimensional systems.} Every finite-dimensional von Neumann algebra is equal to a direct sum of full algebras of linear bounded operators on a finite-dimensional Hilbert space. In the following we will treat every finite-dimensional system as a full algebra of linear bounded operators on a finite-dimensional Hilbert space: $\cM=\cB(\mathbb{C}^{n})$.


\section{Min- and Max-Entropy}

Here we discuss the conditional min- and max-entropy (Section~\ref{sec:cond_minmax}), the corresponding smoothed versions (Sections~\ref{sec:purified}~--~\ref{sec:properties_smooth}), as well as the min- and max-relative entropy (Section~\ref{sec:relative_minmax}).


\subsection{Conditional Min- and Max-Entropy}\label{sec:cond_minmax}

In order to define the conditional entropy of $A$ given the side information $B$, we need a trace on the $A$-system. For this reason we restrict ourselves in this section to von Neumann algebras of the form $\cM_{AB} = \cB(\cH_A)\otimes\cM_{B}$ (see Section~\ref{sec:setup} for a discussion about such systems). We note that the $B$-system is fully general.

\begin{definition}\label{def:Hmin}
Let $\cM_{AB} = \cB(\cH_A)\otimes\cM_{B}$ and $\omega_{AB}\in \cS_{\leq}(\cM_{AB})$. The conditional min-entropy is defined as
\begin{align}
\Hmin{A}{B}_{\omega}:= -\log \inf_{\sigma_B\in \cN^+(\cM_{B})} \Big\{\sigma_B(\idty_B) \,:\,\tau_A \otimes \sigma_B - \omega_{AB} \geq 0 \,\Big\}\,,
\end{align}
where $\tau_A$ denotes the trace on $\cB(\cH_A)$.
\end{definition}

In order to define the conditional max-entropy, we have to make a few comments on the setup. Let \(\pi:\cM_A \otimes \cM_B \to \cB(\cH)\) be a representation of the von Neumann algebra \(\cM_A \otimes \cM_B\) on a Hilbert space \(\cH\). Since \(\cM_A\) is a full algebra, we can decompose the Hilbert space \(\cH = \cH_1 \otimes \cH_2\), such that \(\pi(\cM_A) = \cB(\cH_1) \otimes \idty_{\cH_2}\) with \(\cH_A \simeq \cH_1\). Here \(\pi(\cM_A)\) denotes the von Neumann algebra generated by elements \(\pi(x\otimes \idty)\), \(x \in \cM_A\). By the properties of the von Neumann tensor product, we have that the purifying system is \(\cM_C := \pi(\cM_A \otimes \cM_B)^\prime = \pi(\cM_A)^\prime \otimes \pi(\cM_B)^\prime = \idty \otimes \pi(\cM_B)^\prime\). Any vector \(\ket{\xi} \in \cH\) induces a state on any von Neumann subalgebra of \(\cB(\cH)\), and hence also on the von Neumann algebra \(\cM_{AC} := \pi(\cM_A) \otimes \pi(\cM_B)^\prime = \cB(\cH_A) \otimes \pi(\cM_B)^\prime\). We denote the corresponding state by \(\xi_{AC}\).

For the definition of the conditional max-entropy, we are interested in such a state if \(\ket{\xi}\) is a purification of a state \(\omega_{AB}\) on \(\cM_A \otimes \cM_B\). Especially, we will define the conditional max-entropy as the conditional min-entropy of the state \(\xi_{AC}\) of the system A given the system C. For this to make sense, we first have to show that the definition is independent of the choice of purification.

\begin{lemma}\label{lem:WellDefHmax}
Let $\cM_{AB} =\cB(\cH_A)\otimes\cM_{B}$, $\omega_{AB}\in \cS_{\leq}(\cM_{AB})$, and $(\pi_i,\cK_i, \ket {\xi_i})$ with $i=1,2$ be two purifications of $\omega_{AB}$ with $\pi_i(\M_A\otimes\idty_B)\simeq \M_{A_i}$ and complementary systems $\M_{C_i}$. Then, we have
\begin{align}
\Hmin {A_1}{C_1}_{\omega^1} = \Hmin {A_2}{C_2}_{\omega^2}\,,
\end{align}
where $\omega^i_{A_iC_i}$ are the restricted states corresponding to $\ket {\xi_i}$.
\end{lemma}

\begin{proof}
The conceptual idea for the proof is from~\cite[Lemma 13]{Tomamichel09}. It is straightforward to see that there exists a partial isometry $V:\cK_1\rightarrow \cK_2$ with $\ket{\xi_2}=V\ket{\xi_1}$ and $V\pi_1(a)=\pi_2(a)V$ for all $a\in\M_{AB}$. It follows for all $x\in\M_{A_2}\otimes\M_{C_2} = \cB(\cH_1) \otimes \pi_2(\cM_B)^\prime$ that
\begin{align}\label{pf:eq1:WellDefHmax}
\omega_{A_2C_2}^2(x)=\Scp {\xi_2}{x\xi_2}=\Scp {\xi_1}{V^*xV\xi_1}=\omega_{A_1C_1}^1(V^*xV)\,,
\end{align}
where we used in the last equality that $V^*xV\in\M_{A_1}\otimes\M_{C_1}$. This follows from the fact that $(\M_{A_i}\otimes\M_{C_i})'= \idty\otimes\pi_i(\M_{B})^{\prime\prime} =\pi_i(\M_B)$ and that for all $y\in\M_B$
\begin{align}
V^*xV\pi_1(y)= V^*x\pi_2(y)V= V^*\pi_2(y)xV =  \pi_1(y)V^*xV \,.
\end{align}
From~\eqref{pf:eq1:WellDefHmax} we get that $\omega^1_{A_1C_1}\leq\tau_{A_1}\otimes \sigma_{C_1}$ implies $\omega^2_{A_2C_2}\leq V^*(\tau_{A_1}\otimes \sigma_{C_1})V$ with $V^*(\tau_{A_1}\otimes \sigma_{C_1})V(x)=\tau_{A_1}\otimes \sigma_{C_1}(V^*xV)$. Note that $\cM_{C_2}$ is mapped by $V$ into $\cM_{C_1}$, that is, for any $c\in\cM_{C_2}$ we find that $V^* c V$ lies in $\cM_{C_1}$. This follows from
\begin{align}  
\Scp{\phi}{V^*(\idty_{A_2}\otimes c) V \pi_1(x) \psi}=\Scp{\phi}{V^*(\idty_{A_2}\otimes c) \pi_2(x) V \psi}& = \Scp{\phi}{V^*\pi_2(x)(\idty_{A_2}\otimes c)  V \psi} \\
&= \Scp{\phi}{\pi_1(x)V^*(\idty_{A_2}\otimes c)  V \psi}
\end{align}
for any $x\in\cM_{AB}$ and $\phi,\psi\in\cK_2$. This then implies that for $x\in\cM_A$,
\begin{align}
\tau_{A_1}\otimes \sigma_{C_1}(V^*(\pi_2(x)\otimes c) V)= \tau_{A_1}(\pi_1(x))\sigma_{C_1}(V^*(\idty\otimes c) V)
\end{align} 
factorizes and can therefore be written in the form $\tau_{A_2}\otimes \sigma_{C_2}$, where $\sigma_{C_2}(c)=\sigma_{C_1}(V^*\idty\otimes c V)$. This follows since the tensor product weight is uniquely determined by its value on elementary tensors. With $\sigma_{C_2}(\idty)\leq\sigma_{C_1}(\idty)$ we can conclude that $\Hmin {A_1}{C_1}_{\omega^1} \leq \Hmin {A_2}{C_2}_{\omega^2}$. Since the argument is symmetric, we get equality.
\end{proof}

With this result at hand, we can use the definition of a purification on von Neumann algebras (Definition~\ref{def:purification}) to define the conditional max-entropy as the dual quantity of the conditional min-entropy~\cite[Definition 2]{koenig08}.

\begin{definition}\label{def:Hmax}
Let $\cM_{AB} = \cB(\cH_A)\otimes\cM_{B}$ and $\omega_{AB}\in \cS_{\leq}(\cM_{AB})$. The conditional max-entropy is defined as
\begin{align}
\Hmax{A}{B}_{\omega}:=-\Hmin {A'}C_{\omega}\,,
\end{align}
with $\omega_{A'B'C}$ an arbitrary purification $(\pi,\cK,\ket\xi)$ of $\omega_{AB}$ with $\M_{A'B'}=\pi(\M_{AB})$ the relevant system, and $\M_C=\pi(\M_{A'B'})'$ the complementary system.
\end{definition}


\subsection{Purified Distance}\label{sec:purified}

The smooth conditional min- and max-entropy emerge from their non-smooth counterparts by a maximization and minimization, respectively, over states close with respect to a suitable distance measure. The choice of the distance measure influences the properties of the smooth entropies crucially. Here we extend the so-called purified distance~\cite{Tomamichel09} to the setting of von Neumann algebras.

Following~\cite{Bures69}, the fidelity between $\omega,\sigma\in\cS(\cM)$ is defined as
\begin{align}
F_{\M}(\omega,\sigma):=\sup_{\pi}|\Scp {\xi^{\pi}_{\omega}} {\xi_{\sigma}^{\pi}}|^2\,,
\end{align}
where the supremum runs over all representations $\pi$ of $\M$ for which purifications $\ket{\xi^{\pi}_{\omega}}$ and $\ket{\xi_{\sigma}^{\pi}}$ of $\omega$ and $\sigma$ exist. We suppress the subscript $\M$ if clear from the context and simply write $F_{\M}(\vert \xi_{\omega} \rangle, \sigma)$ instead of $F_{\M}(\omega,\sigma)$ if $\ket{\xi_\omega}$ is a purification of $\omega$. Various properties are known for the fidelity~\cite{Bures69,Uhlmann76,Alberti-1983}. Among them is the monotonicity under quantum channels $\EE$,
\begin{align}\label{Fidelity-Monotonicity}
F(\omega,\sigma)\leq F(\EE_*(\omega),\EE_*(\sigma))\,,
\end{align}
and moreover that $F_{\M}(\omega,\sigma)\leq F_{\cN}(\omega,\sigma)$ for von Neumann algebras $\cN\subset\M$. Furthermore, we can fix a particular representation $\pi$ on $\H$ in which $\omega,\sigma$ admit vector states $\ket{\xi_{\omega}},\ket{\xi_{\sigma}}\in\H$, and get
\begin{align}\label{eq:maxalbertiformfidelity}
F(\omega,\sigma)=\sup_{U\in \pi(\M)'} |\Scp {\xi_{\omega}}{U\xi_{\sigma}}|^2\,,
\end{align}
where the supremum is taken over all elements $U$ in $\pi(\M)'$ with $\Vert U \Vert \leq 1$~\cite{Alberti-1983}.

Following work for finite-dimensional spaces~\cite[Definition 2]{Tomamichel09}, we generalize the fidelity to sets of subnormalized states. We first introduce the concept of a projective embedding. Let $\M$ and $\cN$ be von Neumann algebras. We say that $\N$ admits a projective embedding of $\M$, denoted by $\M\pemb\cN$, if there exists a projector $p\in\cN$ such that $p\cN p$ is isomorphic to $\M$. (Note that if $\M\subset\cB(\H)$ and $V:\H \rightarrow \H'$ is an isometry, it follows that $\M\pemb\cB(\H')$ with the projector $p=VV^*$.) This is equivalent to the existence of a projector $p$ in $\cN$ and a faithful representation $\pi$ of $\M$ into $\N$ such that $\pi(\M)= (\idty-p)\oplus p\cN p$. Given $\omega\in\cS_{\leq}(\M)$ and $\M \pemb \N$ with $\M \cong p\N p$, there exists an extended state $\bar\omega\in\cS(\N)$ such that $\bar\omega(pxp)=\omega(x)$ for $x\in\N$, where we identified $\M$ and $p\N p$. (Choose for instance $\bar\omega(x)=\omega(pxp) + \sigma((\idty-p)x(\idty-p))$ with $\sigma\in\cS_{\leq}(\cN)$ such that $\sigma(\idty-p)=1-\omega(p)$.) Hence, we can interpret subnormalized states as post-measurement states conditioned on certain outcomes.

\begin{definition}\label{def:genFidelity}
Let $\omega,\sigma\in\Ss(\M)$. The generalized fidelity between $\sigma$ and $\omega$ is defined as
\begin{align}\label{def:eq1:genFidelity}
\cF_{\M}(\omega,\sigma):= \sup_{\cM \curvearrowright\cN}\sup_{\bar\omega , \bar\sigma\in\S(\cN)}F_{\cN}(\bar\sigma,\bar\omega)\,,
\end{align}
where the second supremum is taken over all extended normalized states on $\N$ such that $\bar\omega(p \cdot p)$ on $p\N p \cong \M$ corresponds to $\omega$, and similarly for $\bar\sigma$.
\end{definition}

Due to $\M \pemb \M\oplus \mathbb C$, the generalized fidelity can be simplified.

\begin{lemma}\label{lem:statesGenFidelity}
Let $\omega,\sigma\in\Ss(\M)$. Then, we have
\begin{align}
\cF_{\M}(\omega,\sigma)=F_{\hat\cM}(\hat\omega,\hat\sigma)=\left(\sqrt{F_{\M}(\omega,\sigma)}+\sqrt{(1-\omega(\idty))}\sqrt{(1-\sigma(\idty))}\right)^2\,,
\end{align}
where $\hat\M = \M \oplus \mathbb C$, $\hat\omega=\omega\oplus (1-\omega(\idty))$, and $\hat\sigma=\sigma\oplus (1-\sigma(\idty))$.
\end{lemma}

\begin{proof}
Let $\cN$ be such that $\M\pemb\cN$ with $p$ the projector such that $\M \cong p\cN p$. Furthermore, let $\bar\omega,\bar\sigma$ be extensions of $\omega,\sigma$ on $\cN$ satisfying the required properties. We have that $F_{\cN}(\bar\omega,\bar\sigma)=\sup \vert\Scp{\xi_{\bar\omega}^{\pi}}{\xi_{\bar\sigma}^{\pi}}\vert^2$, where the supremum runs over representations of $\cN$. Note that all such representations $\pi$ are also representations of $\M$, that $\xi_{\omega}^{\pi}=\pi(p)\xi_{\bar\omega}^{\pi}$ is a purification of $\omega$, and that the same also holds for $\xi_{\sigma}^{\pi}=\pi(p)\xi_{\bar\sigma}^{\pi}$. We can use the Cauchy-Schwarz inequality to compute
\begin{align}
\vert\Scp{\xi_{\bar\omega}^{\pi}}{\xi_{\bar\sigma}^{\pi}}\vert= \vert\Scp{\xi_{\omega}^{\pi}}{\xi_{\sigma}^{\pi}}\vert + \vert\Scp{(\idty-p)\xi_{\bar\omega}^{\pi}}{(\idty-p)\xi_{\bar\sigma}^{\pi}}\leq \vert\Scp{\xi_{\omega}^{\pi}}{\xi_{\sigma}^{\pi}}\vert + \sqrt{(1-\omega(\idty))(1-\sigma(\idty))}\,.
\end{align}
Since this holds for all $\pi$, we have that
\begin{align}
\F_{\cN}(\bar\omega,\bar\sigma)\leq\left(\sqrt{F_{\M}(\omega,\sigma)}+\sqrt{(1-\omega(\idty))}\sqrt{(1-\sigma(\idty))}\right)^2
\end{align}
for all $\cN$ such that $\M\pemb\cN$ and all suitable $\bar\omega,\bar\sigma$ on $\cN$. Hence, we get
\begin{align}
\cF_{\M}(\omega,\sigma)\leq\left(\sqrt{F_{\M}(\omega,\sigma)}+\sqrt{(1-\omega(\idty))}\sqrt{(1-\sigma(\idty))}\right)^2\,.
\end{align}
Finally it is easy to check that the specific choice $\hat\M$ together with $\hat\omega$ and $\hat\sigma$ achieves equality.
\end{proof}

The purified distance is then defined in the same way as for finite-dimensional spaces~\cite[Definition 4]{Tomamichel09}.

\begin{definition}\label{def:purified_distance}
Let $\omega,\sigma\in\Ss(\M)$. The purified distance between $\rho$ and $\sigma$ is defined as
\begin{align}\label{def:eq1:purified_distance}
\cP_{\M}(\omega,\sigma):=\sqrt{1-\cF_{\M}(\omega,\sigma)}\,.
\end{align}
\end{definition}

The name purified distance comes from the finite-dimensional case, where the purified distance between two states corresponds to the minimal $l_{1}$-distance between purifications of these states. It is straightforward to see that the same result also holds in the von Neumann case, namely,
\begin{align}
\cP_{\cM}(\omega,\sigma)=\frac{1}{2}\inf_{\pi}\|\ket{\xi^{\pi}_{\omega}}\bra{\xi^{\pi}_{\omega}}-\ket{\xi^{\pi}_{\sigma}}\bra{\xi^{\pi}_{\sigma}}\|_{1}\,,
\end{align}
where the infimum runs over all representations of $\M$ in which $\omega$ and $\sigma$ have a vector representation denoted by $\ket{\xi^{\pi}_{\omega}}$ and $\ket{\xi_{\sigma}^{\pi}}$, respectively.

As for the fidelity, we often omit the indication of the von Neumann algebra and moreover write $\cP_{\M}(\omega,\sigma)=\cP_{\M}(\ket \xi , \sigma)$ if $\ket \xi$ is a purification of $\omega$. A detailed discussion of the properties of the purified distance can be found in~\cite{Tomamichel09}. (Although this discussion is restricted to systems described by finite-dimensional spaces, many of the properties follow in the same way for general systems.) It is for instance easy to see that the purified distance defines a metric on $\Ss(\M)$ that is equivalent to the norm distance on $\cN(\M)$,
\begin{align}\label{eq,PurDistEquivNorm}
\sqrt{\Vert \omega -\sigma \Vert + \vert \omega(\idty)-\sigma(\idty)\vert}\geq\cP_{\M}(\sigma,\omega)\geq \frac{1}{2}\Big(\Vert \omega -\sigma \Vert + \vert \omega(\idty)-\sigma(\idty)\vert\Big)\,.
\end{align}
Furthermore, the purified distance is monotone under completely positive contractions.


\subsection{Smooth Conditional Min- and Max-Entropy}\label{sec:smoothcond}

As in Section~\ref{sec:cond_minmax}, we restrict ourselves to von Neumann algebras of the form $\cM_{AB} = \cB(\cH_A)\otimes\cM_{B}$. The smooth entropies are defined using an $\epsilon$-ball with respect to the purified distance,
\begin{align}
\cB^{\epsilon}_{\M}(\omega):=\Big\{\sigma\in\Ss(\M):\cP_{\M}(\omega,\sigma)\leq\epsilon\Big\}\,.
\end{align}
The set $\cB^{\epsilon}_{\M}(\omega)$ is referred to as the smoothing set and $\epsilon$ is called the smoothing parameter.

\begin{definition}\label{def:smooth_entropy}
Let $\cM_{AB} = \BB(\H_A)\otimes\cM_{B}$, $\omega_{AB}\in \cS_{\leq}(\cM_{AB})$, and $\epsilon\geq0$. The $\epsilon$-smooth conditional min-entropy is defined as
\begin{align}\label{def:eq1:smooth_entropy}
\SHmin{A}{B}_{\omega}:=\sup_{\sigma_{AB}\in\Ball{\cM}{\omega_{AB}}}\Hmin AB_{\sigma}\,.
\end{align}
\end{definition}

Since the purified distance defines a metric on $\Ss(\M_{AB})$, we retrieve the conditional min-entropy for $\epsilon=0$. In order to define the smooth conditional max-entropy as the dual quantity of the smooth conditional min-entropy, we again have to make sure that everything is independent of the choice of the purification.

\begin{lemma}\label{lem:InvarianceSmoothHmin}
Let $\cM_{AB} = \BB(\H_A)\otimes\cM_{B}$, $\omega_{AB}\in\Ss(\M_{AB})$, and $(\pi_i,\cK_i, \ket {\xi_i})$ for $i=1,2$ two purifications of $\omega_{AB}$ with $\M_{A_i}=\pi_i(\M_A)$ and complementary systems $\M_{C_i}$, and $\epsilon\geq0$. Then, we have
\begin{align}\label{lem:eq1:InvarianceSmoothHmin}
\SHmin {A_1}{C_1}_{\omega^1} = \SHmin {A_2}{C_2}_{\omega^2}\,,
\end{align}
where $\omega^i_{A_iC_i}$ are the restricted states corresponding to $\ket {\xi_i}$.
\end{lemma}

\begin{proof}
We observe that due to the symmetry of~\eqref{lem:eq1:InvarianceSmoothHmin} it is enough to show inequality in one direction. It is straightforward to see that there exists a partial isometry $V:\cK_1\rightarrow \cK_2$ with $\ket{\xi_2}=V\ket{\xi_1}$ and $V\pi_1(a)=\pi_2(a)V$ for all $a\in\M_{AB}$. Furthermore, it follows from the proof of Lemma \ref{lem:WellDefHmax} that for all $\sigma_{A_1C_1} \in \Ss(\M_{A_1C_1})$ the subnormalized state $V^*\sigma_{A_1C_1}V(x)=\sigma_{A_1C_1}(V^* xV)$ on $\M_{A_2C_2}$ satisfies $\Hmin {A_1}{C_1}_\sigma\leq \Hmin {A_2}{C_2}_{V^*\sigma V}$ and $V^*\omega^1_{A_1C_1}V=\omega_{A_2C_2}^2$. We have
\begin{align}
\SHmin {A_1}{C_1}_{\omega^1}=\sup_{\sigma_{A_1C_1}\in\Ball {}{\omega_{A_1C_1}^1}} \Hmin {A_1}{C_1}_{\omega}\leq \sup_{\sigma_{A_1C_1}\in\Ball {}{\omega_{A_1C_1}^1}} \Hmin {A_2}{C_2}_{V^*\sigma V}\,,
\end{align}
and we are left to prove $V^*\sigma_{A_1C_1}V\in\Ball {}{\omega_{A_2C_2}^2}$ for all $\sigma_{A_1C_1}\in\Ball {}{\omega_{A_1C_1}^1}$. This is equivalent to
\begin{align}
\cF(\omega_{A_1C_1},\sigma_{A_1C_1})\leq\cF(V^*\omega_{A_1C_1}V,V^*\sigma_{A_1C_1}V)\,.
\end{align}
Let $p=VV^*$ be the projector onto the image of $V$. Note that
\begin{align}
V^*\omega_{A_1C_1}V(p) =V^*\omega_{A_1C_1}V(\idty)\quad\mathrm{and}\quad V^*\sigma_{A_1C_1}V(p) =V^*\sigma_{A_1C_1}V(\idty)
\end{align}
holds by construction. Since $p\M_{A_2C_2}p$ is a von Neumann algebra and using Definition~\ref{def:genFidelity}, we find that
\begin{align}
\cF_{\M_{A_2C_2}}(V^*\omega_{A_1C_1}V,V^*\sigma_{A_1C_1}V) & = \cF_{p\M_{A_2C_2}p}(V^*\omega_{A_1C_1}V,V^*\sigma_{A_1C_1}V)\\
&=\sup_{p\M_{A_2C_2}p\ \pemb \hat\cN}\sup_{\bar\omega,\bar\sigma} F_{\cN}(\bar\omega,\bar\sigma)\\
& \geq F_{\hat\M_{A_1C_1}}(\hat\omega^1_{A_1C_1},\hat\sigma_{A_1C_1})\\
&= \cF_{\M_{A_1C_1}}(\omega^1_{A_1C_1},\sigma_{A_1C_1})\,,
\end{align}
where $\hat\M_{A_1C_1},\hat\omega^1_{A_1C_1},\hat\sigma_{A_1C_1}$ are as in Lemma \ref{lem:statesGenFidelity}, the inequality follows from $p\M_{A_2C_2}p \ \pemb\hat\M_{A_1C_1}$ via the isometry $V\oplus 1$, and $\hat\omega^1_{A_1C_1},\hat\sigma_{A_1C_1}$ are extensions of $ V^*\omega_{A_1C_1}V,V^*\sigma_{A_1C_1}V$ in accordance with (\ref{def:eq1:genFidelity}).
\end{proof}

We are now ready to define the smooth conditional max-entropy.

\begin{definition}\label{def:SmoothHmax}
Let $\cM_{AB} = \cB(\cH_A)\otimes\cM_{B}$, $\omega_{AB}\in \cS_{\leq}(\cM_{AB})$, and $\eps\geq0$. The $\epsilon$-smooth conditional max-entropy is defined as
\begin{align}
\SHmax{A}{B}_{\omega}:=-\SHmin {A'}C_{\omega}\,,
\end{align}
with $\omega_{A'B'C}$ an arbitrary purification $(\pi,\cK,\ket\xi)$ of $\omega_{AB}$ with $\M_{A'B'}=\pi(\M_{AB})$ the relevant system, and $\M_C=\pi(\M_{A'B'})'$ the complementary system.
\end{definition}

Lemma~\ref{lem:InvarianceSmoothHmin} ensures that the definition of the smooth conditional max-entropy is independent of the purification. Another possible definition of the smooth conditional max-entropy would have been to smooth the conditional max-entropy (in analogy to the definition of the smooth conditional min-entropy). However, as for finite-dimensional spaces, the two approaches are equivalent~\cite[Lemma 16]{Tomamichel09}. In order to show this, we first need the following lemma.

\begin{lemma}\label{lem:helpdualsmooth}
Let \(\cM \subset \cB(\cH)\), \(\ket{\xi} \in \cH\) be a vector inducing a state \(\omega\) on \(\cM\), and \(\sigma\in\cS(\cM)\) with \(\cP_{\M}(\omega,\sigma) < \infty\). Then, there exists a vector \(\ket{\gamma} \in \cH\) such that \(\Scp{\gamma}{x \gamma} \leq \sigma(x)\;\forall x \in \cM_+ \), and moreover
\begin{align}
\cP_{\M}(\omega,\sigma) = \cP_{\cB(\cH)}(\ket{\xi},\ket{\gamma})\,.
\end{align}
\end{lemma}
  
\begin{proof}
Let \(\pi,\cK\) be a tuple of a Hilbert space \(\cK\) and a representation of \(\cM\) on \(\cK\) such that there exists purifying vectors \(\ket{\tilde{\xi}} \in \cK\) for \(\omega\) as well as \(\ket{\chi} \in \cK\) for \(\sigma\). This can for example be achieved by a GNS construction with respect to the positive functional \(\omega + \sigma\) (since we both have \(\omega \leq \omega + \sigma\) as well as \(\sigma \leq \omega + \sigma\)). It follows that there exists a partial isometry \(V:\cH \to \cK\) with \(V\ket{\xi} = \ket{\tilde{\xi}}\) satisfying \(\pi(x)V = V x\), \(x \in \cM\). We then set \(\ket{\gamma} = V^* U \ket{\chi}\), where \(U \in \pi(\cM)^\prime \) with $\Vert U\Vert \leq 1$ is taken such that \(F(\sigma,\omega) = \Scp{\chi}{U^* V\xi}\). We find for any $x\in\cM_+$ that
\begin{align}
\Scp{\gamma}{x\,\gamma} = \Scp{\chi}{U^* V x\, V^* U \chi} = \Scp{\chi}{ \, \pi(x)^{1/2} U^* V V^* U \pi(x)^{1/2}  \chi} \leq \Scp{\chi}{\pi(x) \,\chi} = \sigma(x)\,,
\end{align}
as well as \(\cP_{\M}(\omega,\sigma) = \cP_{\cB(\cH)}(\ket{\xi},\ket{\gamma})\).
\end{proof}

We can now show that the smooth conditional max-entropy can be written as an optimization over conditional max-entropies.

\begin{proposition}\label{dualitySmooth}
Let $\cM_{AB} = \BB(\H_A)\otimes\cM_{B}$, $\omega_{AB}\in\Ss(\M_{AB})$, and $\epsilon\geq0$. Then, we have
\begin{align}\label{eq:dualitySmooth}
\SHmax AB_{\omega} = \inf_{\sigma_{AB}\in\Ball{}{\omega_{AB}}}\Hmax AB_{\sigma}\,.
\end{align}
\end{proposition}

\begin{proof}
Let $(\pi,\cK,\ket\xi )$ be an arbitrary purification of $\omega_{AB}$ with complementary system $\M_{C}=\pi(\M_{AB})'$. Because of the independence of the smooth conditional min-entropy of a particular purification (Lemma~\ref{lem:InvarianceSmoothHmin}), we can assume that $\pi$ together with $\cK$ is a standard form of $\M$. Thus, each state in $\M_{AB}$ admits a purification in $\cK$. According to the definition of the smooth entropies we have to show
\begin{align}\label{pf:eq1:dualitySmooth}
\sup_{\sigma_{AB}\in \Ball {\cM}{\omega_{AB}}}\Hmin {A}{C}_{\ket{\xi_{\sigma}}}= \sup_{\eta_{AC}\in\Ball {}{\omega_{AC}}} \Hmin AC_{\eta}\,,
\end{align}
where $\ket{\xi_{\sigma}}\in\cK$ is a purification of $\sigma_{AB}$. Since we know that the conditional min-entropy does not depend on the particular choice of the purification $\ket{\xi_{\sigma}}$ (Lemma \ref{lem:WellDefHmax}), we can choose $\ket{\xi_{\sigma}}$ such that $\cF_{\M_{AB}}(\omega_{AB},\sigma_{AB})=\cF_{\cB(\H)}(\ket{\xi},\ket{\xi_{\sigma}})$, and thus
\begin{align}
\cP_{\M_{AB}}(\omega_{AB},\sigma_{AB})=\cP_{\cB(\H)}(\ket\xi,\ket{\xi_{\sigma}})\geq \cP_{\M_{AC}}(\ket\xi,\ket{\xi_{\sigma}})\,,
\end{align}
from which $\leq$ in~\eqref{pf:eq1:dualitySmooth} follows.

For the other direction, let $(\pi,\cH,\ket{\xi} )$ be a purification of \(\omega_{AC}\), and for any element \(\eta_{AC} \in \Ball {}{\omega_{AC}}\) let \(\ket{\gamma(\eta_{AC})} \in \cH\) be the vector obtained from applying Lemma~\ref{lem:helpdualsmooth} to \(\eta_{AC}\), which in turn induces a subnormalized state \(\gamma_{AC}(\eta_{AC})\) on \(\cM_{AC}\). Since \(\gamma_{AC} \leq \sigma_{AC}\), it follows that
\begin{align}
  \sup_{\eta_{AC}\in\Ball {}{\omega_{AC}}} \Hmin AC_{\eta} \leq \sup_{\gamma_{AC}(\eta_{AC})\,:\,\eta_{AC}\in\Ball {}{\omega_{AC}}} \Hmin AC_{\gamma_{AC}} \,.
\end{align}
But \(\gamma_{AC}\) originates from a vector \(\ket{\gamma}\) such that \(\cP_{\cB(\cH)}(\ket{\xi},\ket{\gamma}) \leq \eps\) and hence we find
\begin{align}
\sup_{\eta_{AC}\in\Ball {}{\omega_{AC}}} \Hmin AC_{\eta} &\leq \sup_{\ket{\gamma} \in \cH\,:\,\cP_{\cB(\cH)}(\ket{\xi},\ket{\gamma}) \leq \eps} \Hmin AC_{\ket{\gamma}}\\
&\leq \sup_{\ket{\gamma} \in \cH\,:\,\cP_{\cM_{AB}} (\xi_{AB},\gamma_{AB}) \leq \eps} \Hmin AC_{\ket{\gamma}}\,,
\end{align}
where the last step follows from the fact that the purified distance is monotone. The assertion follows since \(\xi_{AB} = \omega_{AB}\).
\end{proof}


\subsection{Properties of Smooth Entropies}\label{sec:properties_smooth}

\textbf{Data Processing.} The principle that local operations on the quantum side information $B$ can never increase the knowledge about the $A$-system is expressed by the data-processing inequality.

\begin{proposition}\label{prop:processing}
Let $\omega_{AB}\in\cS_{\leq}(\BB(\H_A)\otimes\cM_{B})$, $\EE: \M_C \rightarrow \M_B$ be a quantum channel, and $\epsilon\geq0$. Then, we have
\begin{align}\label{eq:DataProcessingIneq}
\SHmin {A}{B}_{\omega} \leq \SHmin {A}{C}_{\mathcal{I}_A\otimes\EE_*(\omega)}\quad\text{as well as}\quad\SHmax {A}{B}_{\omega} \leq \SHmax {A}{C}_{\mathcal{I}_A\otimes\EE_*(\omega)}\,,
\end{align}
where $\mathcal{I}_{A}:\cM_A\to\cM_A$ denotes the identity map. Moreover, we have that access to partial information can only increase the entropies, that is,
\begin{align}
\M_C \subset \M_B\quad\Rightarrow\quad\SHmin {A}{B}_{\omega} \leq \SHmin {A}{C}_{\omega}\quad\mathrm{and}\quad\SHmax {A}{B}_{\omega} \leq \SHmax {A}{C}_{\omega}\,.
\end{align}
\end{proposition}

The proof of the first statement is obtained by adapting the one for systems described by finite-dimensional spaces~\cite[Theorem 18]{Tomamichel09}. The second statement is obtained from the fact that by restricting the state to a subalgebra, the ordering relation in the definition of the min-entropy~\eqref{def:Hmin} is only tested on fewer positive elements such that the infimum in~\eqref{def:Hmin} is taken over a larger set of states. This then leads to a larger min-entropy and thus, to a larger smooth min-entropy.\\

\textbf{Bounds.} Here we would like to study when the smooth conditional min- and max-entropy are finite.

\begin{proposition}\label{prop:dimension2}
Let $\cM_{AB}=\cB(\cH_A)\otimes\cM_{B}$ and $\omega_{AB}\in\cS(\cM_{AB})$. Then, we have
\begin{align}
\Hmin{A}{B}_{\omega}<\infty\quad\mathrm{and}\quad \Hmax{A}{B}_{\omega}>-\infty\,.
\end{align}
\end{proposition}

\begin{proof}
The first inequality follows from the data processing inequality (Proposition~\ref{prop:processing}) and the corresponding statement for the unconditional min-entropy (which we will show in Proposition~\ref{prop:bounds} for a more general setup). The second inequality follows from the duality of the conditional min- and max-entropy (Definition~\ref{def:Hmax}) and the first inequality.
\end{proof}

Note that the conditional min-entropy can become minus infinity and the conditional max-entropy can become plus infinity~\cite[Lemma 1]{Furrer10}. However, the smooth conditional min- and max-entropy with smoothing parameter $\epsilon >0$ are always finite.

\begin{proposition}
Let $\cM_{AB}=\cB(\cH)\otimes\cM_{B}$, $\omega_{AB}\in\cS(\cM_{AB})$, and $\eps>0$. Then, we have
\begin{align}
-\infty<\SHmin{A}{B}_{\omega}<\infty\quad\mathrm{and}\quad-\infty<\SHmax{A}{B}_{\omega}<\infty\,.
\end{align}
\end{proposition}

\begin{proof}
The inequalities $\SHmin{A}{B}_{\omega}<\infty$ and $\SHmax{A}{B}_{\omega}>-\infty$ follow by the corresponding statements for the non-smooth entropies (Proposition~\ref{prop:dimension2}). The other two inequalities follow from applying~\cite[Lemma 2]{Furrer10} together with the data processing inequality (Proposition~\ref{prop:processing}). Namely, let \((\pi,\cH,\ket{\xi})\) be a purification of \(\omega\). Since \(\M_A\) is a full algebra, we can find a decomposition \(\cH = \cH_1 \otimes \cH_2\), \(\cH_1 \simeq \cH_A\) and \(\M_{AB} = \cB(\cH_A) \otimes \pi(\M_B) \subset \cB(\cH_A \otimes \cH_2)\). The vector \(\ket{\xi}\) then induces a normal state on \(\cB(\cH_A) \otimes \cB(\cH_2)\) and we can apply~\cite[Lemma 2]{Furrer10}, followed by the restriction onto the subalgebra \(\pi(\M_B) \subset \cB(\cH_2)\).
\end{proof}


\subsection{Min- and Max-Relative Entropy}\label{sec:relative_minmax}

Instead of conditional entropies we can also define a min- and max-version of relative entropy (as noticed in~\cite[Definition 1]{datta08} for finite-dimensional spaces). This will also allow us to define the (unconditional) min- and max-entropy on von Neumann algebras.

\begin{definition}
Let $\omega,\sigma\in\cN^{+}(\cM)$. The max-relative entropy of $\omega$ with respect to $\sigma$ is defined as
\begin{align}
\Dmax{\omega}{\sigma}:=\inf\left\{\mu\in\mathbb{R}:\omega \leq 2^{\mu} \cdot \sigma\right\}\,,
\end{align}
where the infimum of the empty set is defined to be $\infty$. The min-relative entropy of $\omega$ with respect to $\sigma$ is defined as
\begin{align}
\Dmin{\omega}{\sigma}:=-\log F(\omega,\sigma)\,.
\end{align}
\end{definition}

We have the following ordering relation.

\begin{proposition}\label{prop:relative_ordering}
Let \(\omega \in \cS(\cM)\) and $\sigma\in\cN^{+}(\cM)$. Then, we have
\begin{align}
\Dmin{\omega}{\sigma}\leq \Dmax{\omega}{\sigma}\,.
\end{align}
\end{proposition}

\begin{proof}
If there exists no finite constant $c$ such that \(\omega \leq c\cdot\sigma\), then \(\Dmax{\omega}{\sigma} = \infty\), and there is nothing to prove. So let us suppose the opposite, and let \((\pi,\cK,\ket{\xi})\) be a tuple of a Hilbert space and a representation on it such that there exists purifying vectors \(\ket{\xi} \in \mathcal{K}\) for \(\omega\) as well as \(\ket{\chi} \in \mathcal{K}\) for \(\sigma\). Let \(\mu \in \mathbb{R}\) such that \(\omega \leq 2^\mu\cdot\sigma\). By the non-commutative Radon-Nikodym theorem~\cite[Chapter VII.2]{Takesaki1} there exists an element \(h_\omega \in \pi(\cM)^\prime\) such that \(h_\omega \ket{\chi} = \ket{\xi}\) as well as \(\norm{h_\omega}^2 \leq 2^\mu\). Using the property~\eqref{eq:maxalbertiformfidelity} of the fidelity, we find
\begin{align}
2^\mu\cdot F(\omega,\sigma) \geq 2^\mu \left|\Scp{\xi}{\left(2^{-\mu/2} h_\omega \right) \chi} \right|^2 = \Scp{\xi}{\xi} = 1 \,.
\end{align}
Taking logarithms proves the assertion.
\end{proof}

The following proposition shows that the min- and max-relative entropy are monotone under quantum channels. The proof of the first statement follows by definition, the proof of the second follows from the monotonicity of the fidelity~\eqref{Fidelity-Monotonicity}.

\begin{proposition}\label{prop:important}
Let $\omega,\sigma\in\cS(\cM)$. Then, we have for any quantum channel $\cE$,
\begin{align}
\Dmax{\omega}{\sigma}\geq \Dmax{\cE_*(\omega)}{\cE_*(\sigma)}\quad\mathrm{and}\quad\Dmin{\omega}{\sigma}\geq \Dmin{\cE_*(\omega)}{\cE_*(\sigma)}\,.
\end{align}
\end{proposition}

\textbf{Min- and Max-Entropy.} In the case where the system is given by a full algebra $\cB(\cH)$, the unconditional min- and max-entropy are simply obtained from Definition~\ref{def:Hmin} and Definition~\ref{def:Hmax} (see also Proposition~\ref{thm:max}) with trivial quantum side information. The extension to arbitrary systems can be done similarly as for the von Neumann entropy~\cite[Chapter II.6]{Petz_Book}. 

\begin{definition}\label{def:unconditional}
Let $\omega_{A}\in\cS(\cM_{A})$. The min-entropy is defined as
\begin{align}
\mathrm{H}_{\min}(A)_{\omega}:=-\sup\Big\{D_{\max}(\sigma_{AX}\|\omega_{A}\otimes\tau_{X})|\sigma_{AX}\in\cS(\cM_{A}\otimes\ell^{\infty}_{X}),\sigma_{A}=\omega_{A}\Big\}\,,
\end{align}
where $\tau_{X}(\cdot)$ denotes the trace on the classical system $\ell^{\infty}_{X}$. The max-entropy is defined as
\begin{align}
\mathrm{H}_{\max}(A)_{\omega}:=-\inf\Big\{D_{\min}(\sigma_{AX}\|\omega_{A}\otimes\tau_{X})|\sigma_{AX}\in\cS(\cM_{A}\otimes\ell^{\infty}_{X}),\sigma_{A}=\omega_{A}\Big\}\,.
\end{align}
\end{definition}

\begin{proposition}\label{prop:bounds}
Let $\omega_{A}\in\cS(\cM_{A})$. Then, we have
\begin{align}
0\leq \mathrm{H}_{\min}(A)_{\omega}<\infty\quad\mathrm{and}\quad \mathrm{H}_{\min}(A)_{\omega}\leq \mathrm{H}_{\max}(A)_{\omega}\,.
\end{align}
\end{proposition}

\begin{proof}
The first assertion can be deduced directly from the definition of the min-entropy (Definition~\ref{def:unconditional}). The second assertion follows by the ordering of the min- and max-relative entropy (Proposition~\ref{prop:relative_ordering}).
\end{proof}

Finally, we could also define smoothed versions in the same manner as for the conditional min- and max-entropy.


\section{Applications to Quantum Information Theory}\label{sec:qit}

In the following we restrict ourselves to von Neumann algebras of the form $\cM_{AB} = \cB(\mathbb{C}^n)\otimes\cM_{B}$, that is, the $A$-system is finite-dimensional (see Section~\ref{sec:setup} for a discussion about such systems). We note that the $B$-system is fully general. This setup is well suited for applications in quantum information theory and in particular in quantum cryptography.  (We do not want to make any assumptions about the adversarial system $B$, but our resource, the $A$-system, is finite.)


\subsection{Operational Interpretation of Conditional Min- and Max-Entropy}\label{sec:cond}

\textbf{Optimal entanglement fidelity.} The following proposition generalizes the operational meaning of the conditional min-entropy~\cite[Theorem 2]{koenig08}. (The difference of a square in comparison to~\cite[Theorem 2]{koenig08} is due to the different definition of the fidelity.)

\begin{proposition}\label{cor:HminDualForm}
Let $\cM_{AB} = \cB(\C^n)\otimes\cM_{B}$, $\omega_{AB}\in \cS(\cM_{AB})$, and let $\vert\Phi^n_{AA'}\rangle:=\sum_{i=1}^n\vert \phi_i\rangle\otimes\vert\psi_i\rangle$, where $\{\ket{\phi_i}\}$ and $\{\ket{\psi_i}\}$ are orthonormal bases of $\mathbb{C}^n$. Then, we have 
\begin{align}\label{cor:HminDualForm,eq1}
2^{-\Hmin AB_{\omega}} = \sup_{\EE} F((\mathcal{I}_A\otimes\EE_*)(\omega_{AB}),\vert\Phi^n_{AA'}\rangle)\,,
\end{align}
where the supremum is taken over all quantum channels $\EE: \cB(\C^n)\rightarrow \M_B $.
\end{proposition}

The idea of the proof is that $\Hmin AB_{\omega}$ can be written as the solution of an optimization problem over a subcone of $\cN^+(\M_{AB})$ for which the theory of ordered vector spaces~\cite{PaulsenAOU} applies. For $\omega_{AB}=(\omega_B^{ij})\in\cS_{\leq}(\M_{AB})$, we write
\begin{align}
2^{-\Hmin AB_{\omega}} &= \inf\{\sigma_B(\idty):\tau_A \otimes \sigma_B \geq \omega_{AB} , \, \sigma_B \in \N^+(\M_{B}) \} \\
&= \inf\{f_{\idty}(\sigma_{AB}):\sigma_{AB}\geq\omega_{AB}, \, \sigma_{AB}\in E\}\,,
\end{align}
where $f_{\idty}(\eta_{AB})=\frac{1}{n}\eta_{AB}(\idty)$ for $\eta_{AB}\in \cN(\M_{AB})$, and $E:=\{\tau_A\otimes\eta_B:\eta_B \in \cN^h(\M_{B})\}$ with $\cN^h(\M_{B})$ the set of hermitian functionals on $\M_B$. We have that $E$ is a subspace of $\cN^h(\M_{AB})$ and $f_{\idty}$ defines a positive functional on $\cN^h(\M_{AB})$. The basic ingredient is the following extension result for positive functionals in ordered vector spaces.

\begin{lemma}\cite[Lemma 2.13]{PaulsenAOU}\label{lem:dualform}
Let $V$ be an ordered real vector space with a full cone $V^+$, $E\subset V$ a subspace which majorizes $V^+$, $w \in V\backslash E$, and $f:E\rightarrow \mathbb{R}$ a positive functional on $E$. Then, $f$ admits a positive extension $\tilde{f}$ on $V$ such that
\begin{align}
\tilde f (w) = u_f(w):=\inf\{f(v) \; :\; v\geq w ,\, v\in E\}\,.
\end{align}
Moreover, it holds for all positive functionals $g$ on $V$ with $g|_E=f$, that $g(w)\leq u_f(w)$.
\end{lemma}

If we take $V=\cN^h(\M_{AB})$ with the cone of all positive functionals $V^+=\cN^+(\M_{AB})$ and $E$ as defined above, then $E$ majorizes $V^+$. According to the definition of the predual, the set of all positive functionals on $V$ are given by the positive operators in $\M_{AB}$. Hence, by applying Lemma~\ref{lem:dualform} with $f=f_{\idty}$, we find the following corollary.

\begin{corollary}\label{prop:HminDualForm}
Let $\cM_{AB}=\cB(\C^n)\otimes\cM_{B}$ and $\omega_{AB}=(\omega_B^{ij})\in\Ss(\cM_{AB})$. Then, we have that
\begin{align}\label{prop:HminDualForm,eq1}
2^{-\Hmin AB_{\omega}} = \sup\left\{\sum_{ij} \omega_B^{ij}(M_{ij}) \; :\; (M_{ij})\in (\cB(\C^n) \otimes \M_B)_{+} ,\, \sum_i M_{ii}=\idty\right\}\,.
\end{align}
\end{corollary}

\begin{proof}
The linear functional given by $(M_{ij})$ restricted to $E$ has to be $f_{\idty}$, and thus we have $(\tau_A\otimes \sigma_B)( (M_{ij}))=\sum_i \sigma_B(M_{ii}) =1$ for all $\sigma_{B}\in\cS(\M_{B})$. Since $(\cN(\M_{B}))^*=\M_B$ this implies $\sum_i M_{ii}=\idty$, and the assertion follows.
\end{proof}

The operational form of the conditional min-entropy (Proposition~\ref{cor:HminDualForm}) follows from Corollary~\ref{prop:HminDualForm} by the identification of completely positive maps $\EE: \cB(\C^n)\rightarrow \M_B$ with positive elements $M$ of $\M_{AB}$. Given $M=(M_{ij})\in(\cB(\C^n)\otimes\M_B)_{+}$ with $\sum_i M_{ii}=\idty$, we define the map $\EE$ via $\EE_*(\sigma)=(\sigma(M_{ij})^{ij})$ for $\sigma\in\cN(\M_B)$. $\EE$ is unital because of $\sum_i M_{ii}=\idty$. The states are with respect to the fixed basis in $\cB(\C^n)$ given by $\{\ket{\psi_i}\}$, such that for $A= \sum_{ij}a_{ij}\vert\psi_i\rangle\langle\psi_i\vert$, $(\sigma(M_{ij})^{ij})(A)=\sum_{ij}a_{ij}\sigma(M_{ij})$. It is straightforward to check that $\EE$ is a quantum channel and satisfies
\begin{align}\label{cor:HminDualForm,pf1}
F\big( (\mathcal{I}_A\otimes\EE_*)(\omega_{AB}),\ket{\Phi_{AA'}}\big)=\sum_{ij}\omega_B^{ij}(M_{ij})\,.
\end{align}
The converse is obtained by setting for an arbitrary quantum channel $\EE$, $M_{ij}^{\EE}=\EE(\vert\psi_i\rangle\langle\psi_j\vert)$. It follows directly from complete positivity and unitality that $M^\EE=(M_{ij}^{\EE})$ is positive and $\sum_i M^{\EE}_{ii}=\idty$. The relation~\eqref{cor:HminDualForm,pf1} can be verified straightforwardly.\\

\textbf{Optimal decoupling fidelity.} The following proposition generalizes the operational meaning of the conditional max-entropy~\cite[Theorem 3]{koenig08}.

\begin{proposition}\label{thm:max}
Let $\cM_{AB} = \cB(\C^n)\otimes\cM_{B}$ and $\omega_{AB}\in \cS_{\leq}(\cM_{AB})$. Then, we have that
\begin{align}\label{def:Hmax,dualityRelation}
\Hmax AB_{\omega}= \sup_{\sigma_B\in \cS(\cM_{B})} \log F(\omega_{AB},\tau_A \otimes \sigma_B)\,.
\end{align}
\end{proposition}

\begin{proof}
The statement can be proven in a similar way as for systems described by finite-dimensional spaces~\cite[Theorem 3]{koenig08}. Recall that each state in $\M_{AB}$ can be purified in the standard form, that is, in $\cB(\C^n)\otimes \cB(\C^n)\otimes \M_B^{\phi}$, where $\M_B^{\phi}$ is a standard form of $\M_B$~\cite{schmitt82}. We denote the complementary system by $\M_{A'B'}$ since it consists of a copy of the A-system $\M_{A'}=\cB(\C^n)$ and the commutant $\M_{B'}=(\M_B^{\phi})'$ of the system $B$. Thus, $\M_{ABA'B'}\subset\cB(\cK)$ with $\cK=\mathbb{C}^{2n}\otimes\H_{\phi}$. Let now $\ket{\xi_{\omega}}\in\cK$ be a purification of $\omega_{AB}$ and $\vert\Phi_{AA'}\rangle$ a non-normalized maximally entangled state on $\M_{AA'}$ as in Proposition~\ref{cor:HminDualForm}, thus a purification of $\tau_A$. Then, with $\ket{\eta_{\sigma}}\in\H_{\phi}$ a purification of $\sigma\in\cS(\M_B)$, we find that
\begin{align}
F(\omega_{AB},\tau_A\otimes\sigma_B)= \sup_{U\in\M_{A'B'}} \vert \Scp{\xi_{\omega}}{U(\Phi_{AA'}\otimes\eta_{\sigma})} \vert^2\leq \sup_{U\in\M_{A'B'}} F_{\M_{AA'}}(U\ket{\xi_{\omega}},\ket{\Phi_{AA'}}\otimes\ket{\eta_{\sigma}})\,,
\end{align}
where the supremum is taken over unitaries $U$ in $\M_{A'B'}$. According to Stinespring`s dilation theorem~\cite{stinespring54}, applying a unitary followed by a restriction of the state is a quantum channel, such that the state on $\M_{AA'}$ described by $U\ket{\xi_{\omega}}$ can be obtained by applying a quantum channel $\EE^{U}:\M_{A'} \rightarrow M_{A'B'}$ on $\omega_{AA'B'}$. Hence, together with the operational interpretation of the conditional min-entropy (Proposition~\ref{cor:HminDualForm})
\begin{align}
F(\omega_{AB},\tau_A\otimes\sigma_B)\leq \sup_{U} F_{\M_{AA'}}((\mathcal{I}_A\otimes\EE^U_*)(\omega_{AA'B'}),\ket{\Phi_{AA'}})\leq 2^{-\Hmin A{A'B'}_{\omega}}= 2^{\Hmax AB_{\omega}}\,.
\end{align}
Taking the supremum over all $\sigma_{B}\in\cS(\cM_{B})$, we find inequality in one direction. In order to show the other direction, we note that again by the operational form of the conditional min-entropy (Proposition~\ref{cor:HminDualForm}), there exists for all $\delta>0$ a quantum channel $\EE:\M_{A'}\rightarrow M_{A'B'}$ such that
\begin{align}
2^{\Hmax AB_{\omega}} \leq F((\mathcal{I}_A\otimes\EE_*)(\omega_{AA'B'}),\ket {\Phi_{AA'}})+\delta\,.
\end{align}
Let now $\ket {\xi_{\omega^{\EE}}}$ be a purification of $(\mathcal{I}_{AB}\otimes\EE_*)(\omega_{ABA'B'})$, which can always be found on the extended system $\M_{AA'CBB'}$, where $\M_C=M_{n^2}$. With an arbitrary $\ket {\theta} \in \mathbb{C}^{n^2}\otimes \H_{\phi}$, we obtain
\begin{align}
 F((\mathcal{I}_A\otimes\EE_*)(\omega_{AA'B'}),\ket {\Phi_{AA'}})&= \sup_{U\in\M_{CBB'}} \vert \Scp{\xi_{\omega^{\EE}}}{ U (\Phi_{AA'}\otimes \theta)} \vert^2  \\
& \leq \sup_{U\in\M_{CBB'}} F_{\M_{AB}}(\ket {\xi_{\omega^{\EE}}} , \ket {\Phi_{AA'}}\otimes \ket{ U\theta})\,. 
\end{align}
Since the reduced state of $\ket {\xi_{\omega^{\EE}}}$ on $\M_{AB}$ is $\omega_{AB}$, and there exists for all $\sigma_B\in\cS(\M_B)$ a purification of the form $\ket {U\theta}$ with $U$ unitary in $\M_{CBB'}$, we arrive at
\begin{align}
2^{\Hmax AB_{\omega}} \leq \sup_{\sigma_B\in\cS(\M_B)} F(\omega_{AB},\tau_A\otimes\sigma_{B}) + \delta\,.
\end{align}
Because this holds for any $\delta>0$, we found the inequality in the other direction.
\end{proof}

\textbf{Ordering of entropies.} Given these alternative formulations we find that the conditional min-entropy is never larger than the conditional max-entropy.

\begin{proposition}\label{prop:ordering}
Let $\cM_{AB}=\cB(\mathbb{C}^n)\otimes\cM_{B}$ and $\omega_{AB}\in\cS(\cM_{AB})$. Then, we have
\begin{align}
\Hmin {A}{B}_{\omega}\leq\Hmax {A}{B}_{\omega}\,.
\end{align}
\end{proposition}

This follows directly from ordering of the min- and max-relative entropy (Proposition~\ref{prop:relative_ordering}).


\subsection{Classical Quantum Systems}\label{sec:classical_quantum}

Of particular interest in quantum information theory are correlations between classical and quantum degrees of freedom. A classical system is specified by the property that all observables commute, and is thus described by an abelian von Neumann algebra. We restrict to classical systems over a finite alphabet $X$, given by the bounded complex valued sequences on $X$, $\ell^\infty_X(\mathbb C)$, supplied with the supremum norm. (For general discrete and continuous classical systems see the follow-up work~\cite{furrer13}.) We denote a classical system with alphabet $X$ simply by $X$, and the corresponding algebra by $\ell^\infty_{X}$. A bipartite system consisting of a classical part $X$ and a quantum part $B$ is described by the von Neumann algebra
\begin{align}
\M_{XB}=\ell^\infty_{X}\otimes\M_B\,,
\end{align}
which is isomorphic to the $\cM_B$-valued sequences $\ell^\infty_X(\cM_B)$. States on $\ell^\infty_{X}(M_B)$ are called classical quantum states, and can be written as
\begin{align}
\omega_{XB}=(\omega_B^x)_{x\in X}\;\;\mathrm{with}\;\;\omega_B^x\in\cS_{\leq}(\M_B),\;\;\text{such that}\;\;\omega_{XB}(a)=\sum_x\omega_B^x(a_x)\;\;\forall a=(a_x)\in\M_{XB}\,.
\end{align}
We have the norm
\begin{align}
\Vert (\omega^x) \Vert_{\ell^{1}(\cN(\cM_{B}))} = \sum_{x\in X} \Vert \omega^x \Vert_{\cN(\M_B)}\,.
\end{align}

\textbf{Conditional min-entropy.} Consider the algebra $\cM_{XB}=\ell^\infty_X\otimes\cM_B$ and $\omega_{XB}\in\cS(\cM_{XB})$. This also defines a state \(\omega_{A_{|X|} B}\) on the algebra \(\cM_{A_{|X|} B} = \cB(\mathbb{C}^{|X|})\otimes \cM_B\) by setting \(\omega_B^{xy} = \delta_{xy} \omega_x\). This implies
\begin{align}\label{eq:embedclassintoquant}
\omega_{A_{|X|} B}\left(\sum_{x,y=1}^{|X|} \ket{x}\bra{y} \otimes M_{xy}\right) = \sum_{x} \omega_x(M_{xx}) \,,\quad M=(M_{xy})\in \cB(\C^{|X|}) \otimes \M_B\,,
\end{align}
which is a positive normalized functional. As such, we can compute its conditional min-entropy,
\begin{align}
\Hmin{X}{B}_{\omega}:=\Hmin{A_{|X|}}{B}_{\omega}\,.
\end{align}
It is easily seen that we can also write
\begin{align}
\Hmin{X}{B}_{\omega}=-\log\inf_{\sigma_B\in \cN^+(\cM_{B})} \Big\{\sigma_B(\idty_B) \,:\,\tau_X \otimes \sigma_B - \omega_{XB} \geq 0 \,\Big\}\,,
\end{align}
where $\tau_X$ denotes the trace on $\ell^{\infty}_X$. Using the results from Section~\ref{sec:cond}, we find that the conditional min-entropy of a classical quantum state has an operational interpretation as the probability of correctly guessing the classical register $X$ by making use of the quantum side information $B$~\cite[Theorem 1]{koenig08}.

\begin{corollary}\label{prop:Hmin=guessing}
Let $\cM_{XB}=\ell^\infty_X\otimes\cM_B$ and $\omega_{XB}\in\cS(\cM_{XB})$. Then, we have $\Hmin{X}{B}_{\omega}= -\log p_{\mathrm{guess}}(X|B)_{\omega}$ with
\begin{align}\label{GuessingProb}
p_{\mathrm{guess}}(X|B)_{\omega}= \sup\left\{\sum_{x\in X}\omega_B^x(E_x) \;:\; E_x\in \M_B, \; E_x\geq 0, \; \sum_{x\in X}E_x =\idty\right\}\,,
\end{align}
the guessing probability of the random variable $X$ given the system $B$.
\end{corollary}

The result follows directly from Lemma~\ref{lem:dualform} in analogy to the operational form of the fully quantum conditional min-entropy (Corollary~\ref{prop:HminDualForm}). Moreover, using the embedding as in~\eqref{eq:embedclassintoquant} also allows to define the smooth conditional min-entropy of classical quantum states as
\begin{align}
\SHmin{X}{B}_{\omega}:=\sup_{\sigma_{XB}\in\cB_{\cM_{XB}}^{\eps}(\omega_{XB})}\Hmin{X}{B}_{\sigma}\,.
\end{align}
It follows from the data processing for the smooth conditional min-entropy (Proposition~\ref{prop:processing}) that alternatively we could also smooth over the set $\cB_{\cM_{A_{|X|}B}}^{\eps}(\omega_{A_{|X|}B})$ in the embedding. The proof of this is the same as for finite-dimensional spaces~\cite[Remark 3.2.4]{renner05}.\\

\textbf{Conditional max-entropy.} Again considering a classical quantum system \(\cM_{XB} = \ell_X^\infty \otimes \cM_B\) and a state \(\omega_{XB} \in \cS(\cM_{XB})\), we can use~\eqref{eq:embedclassintoquant} to define a state on the system \(\cM_{A_{|X|} B} =\cB(\mathbb{C}^{|X|})\otimes \cM_B\). This allows us to consider its conditional max-entropy,
\begin{align}
\Hmax{X}{B}_{\omega}:=\Hmax{A_{|X|}}{B}_{\omega}\,.
\end{align}
Using the results from Section~\ref{sec:cond} we find the following characterization.

\begin{corollary}
Let $\cM_{XB}=\ell^\infty_X\otimes\cM_B$ and $\omega_{XB}\in\cS(\cM_{XB})$. Then, we have $\Hmax{X}{B}_{\omega}= \log F_{\mathrm{dec}}(X|B)_{\omega}$ with
\begin{align}
F_{\mathrm{dec}}(X|B)_{\omega}= \sup\left\{\left(\sum_{x\in X}\sqrt{F(\omega_B^x,\sigma_B)}\right)^2\Big|\sigma_B\in\cS(\cM_B)\right\}\,.
\end{align}
\end{corollary}

This follows directly from the characterization of the conditional max-entropy in terms of the optimal decoupling fidelity (Proposition~\ref{thm:max}) together with the fact that the fidelity between two direct sums of states is a sum itself. The smooth conditional max-entropy is then given as
\begin{align}
\SHmax{X}{B}_{\omega}:=\sup_{\sigma_{XB}\in\cB_{\cM_{XB}}^{\eps}(\omega_{XB})}\Hmax{X}{B}_{\sigma}\,,
\end{align}
and again we might alternatively smooth over the set $\cB_{\cM_{A_{|X|}B}}^{\eps}(\omega_{A_{|X|}B})$ in the embedding. For a proof we just follow the arguments for finite-dimensional spaces~\cite[Lemma 3]{renesrenner10}.


\subsection{Entropic Uncertainty Relations with Quantum Side Information}\label{sec:uncertainty}

One of the fundamental principles of quantum mechanics is that for a fixed state the outcome distribution of two measurements described by non-commuting observables cannot be deterministic. A lower bound on the uncertainty inherent by two such measurements is called an uncertainty relation. Since entropies are measures of uncertainty, it is natural to quantify this uncertainty using entropy measures, see the review articles~\cite{Wehner09,Birula10}. Recently it was realized that if one allows to have quantum information about the system in question, the situation qualitatively changes and one has a subtle interplay between uncertainty and entanglement between the observer and the system~\cite{berta10}. This effect is quantified by means of so-called entropic uncertainty relations with quantum side information~\cite{berta10,Tomamichel11,colbeck11,frank12,furrer13}. Besides the fundamental interest, these relations also have manifold applications in quantum cryptography~\cite{tomamichellim11,furrer12,berta12,dupuis13}.\\

\textbf{Measurements.} We start with a tripartite quantum state $\omega_{ABC}\in\cS(\cM_{ABC})$ and two POVMs $\{E_{A}^{x}\}_{x\in X}$ and $\{F_{A}^{y}\}_{y\in Y}$ on system $A$ with finite outcome ranges $X$ and $Y$, respectively. We are then interested in the uncertainty of the outcome distribution of the measurements $\{E_{A}^{x}\}$ and $\{F_{A}^{y}\}$ given the quantum side information $B$ and $C$, respectively. We quantify the uncertainty in terms of the smooth conditional min- and max-entropy.

\begin{proposition}\label{thm:uncert}
Let $\omega_{ABC}\in\cS(\cM_{ABC})$, $\{E_{A}^{x}\}_{x\in X}$ and $\{F_{A}^{y}\}_{y\in Y}$ be POVM's on $\M_{A}$ with finite outcome ranges $X$ and $Y$, and $\epsilon\geq0$. Then, we have that 
\begin{align}\label{thm,eq:uncert}
\SHmin{X}{B}_\omega + \SHmax{Y}{C}_\omega \geq -\log\max_{x,y} \left\|(E_{A}^{x})^{\frac{1}{2}}\cdot(F_{A}^{y})^{\frac{1}{2}}\right\|^{2}\,,
\end{align}
where $\omega_{XB}:=(\omega^{x}_{B})$ with $\omega_{B}^{x}(\cdot):=\omega_{AB}(E_{A}^{x}\cdot)$, and $\omega_{YC}:=(\omega^{y}_{C})$ with $\omega_{C}^{y}(\cdot):=\omega_{AC}(F_{A}^{y}\cdot)$ are classical quantum states on $\ell_X^\infty(\cM_{B})$ and $\ell_Y^\infty(\cM_{C})$, respectively.
\end{proposition}

Note that since we started with a fully general tripartite von Neumann algebra $\cM_{ABC}$, no approximation techniques (as, e.g., from~\cite{Furrer10}) can be applied to just lift the result from finite-dimensions. In the work~\cite{furrer12}, we use the uncertainty relation~\eqref{thm,eq:uncert} to analyze the security of continuous variable quantum key distribution protocols. Moreover, in a follow-up work~\cite{furrer13}, we discuss a non-smooth extension of Proposition~\ref{thm:uncert} for measurements with infinitely many outcomes (discrete and continuous). In the following we will derive Proposition~\ref{thm:uncert} from a more general uncertainty relation that also holds for quantum channels and not only for measurements.\\

\textbf{Quantum Channels.} Here we start with a tripartite quantum state $\omega_{ABC}\in\cS(\cM_{ABC})$ and two quantum channels $\cE:\cM_{E}\rightarrow\cM_{A}$ and $\cG:\cM_{G}\rightarrow\cM_{A}$ with their domains $\M_{E} \cong \cB(\C^{n'})$ and $\cM_{G} \cong \cB(\C^n)$ being matrix algebras. We are then interested in the uncertainties about the quantum systems obtained by the quantum channels $\cE$ and $\cG$ given systems $B$ and $C$ , respectively.

Let us first introduce some notation. By definition the quantum channel $\cE:\cB(\C^n) \to \M_A$ is a completely positive, unital map. As we can always embed $\M_A \subset \cB(\H)$ faithfully for some $\H$, we can apply Stinespring's dilation theorem to $\cE$. There exist a Hilbert space $\H'$, a representation $\pi$ of $\cB(\C^{n'})$ on $\H'$ and an isometry $V : \H \to \H'$, such that $\cE(x) = V^*\, \pi(x) \,V$. If vectors of the form $\pi(x)V\ket{\psi}$, $x\in\cB(\C^{n'})$, $\ket{\psi} \in\mathcal{H}$ are dense in $\H'$, then we call the triple $(V,\H,\pi)$ a minimal Stinespring dilation, which always exists. For such a mininmal dilation, we can choose $\H'$ to be isomorphic to $\C^n \otimes \C^d \otimes \H$ with $1 \leq d \leq n$, and $\pi$ of the form $\pi(x) = x \otimes \idty_{d} \otimes \idty_\H$. From now, on we always assume that the dilation is minimal, unless otherwise stated.

\begin{lemma}\label{thm:mainuncert}
Let $\omega_{ABC}\in\cS(\cM_{ABC})$, $\cE:\cM_{E}\rightarrow\cM_{A}$ and $\cG:\cM_{G}\rightarrow\cM_{A}$ be quantum channels with $\M_{E} \cong \cB(\C^{n'})$ and $\cM_{G} \cong \cB(\C^n)$ being matrix algebras, and $\epsilon\geq0$. If $U:\H \to \C^{n'} \otimes \C^{d'} \otimes \H$ and $V: \H \to \C^{n} \otimes \C^{d} \otimes \H$ denote the isometries corresponding to the minimal Stinespring dilation of $\cE$ and $\cG$, respectively, then we have
\begin{align}
\SHmin{E}{B}_\omega + \SHmax{G}{C}_\omega \geq -\log c(U V^*)\,,
\end{align}
where $\omega_{EB}(x):=\omega_{AB}(\cE(x))$, $\omega_{GC}(y):=\omega_{AC}(\cG(y))$, and
\begin{align}\label{eq:choi}
c(V U^*):= \inf\left\{ c\,>\,0\,:\, c\,\bar{\tr}_{n'} - \cJ_{V^*U} \;\,\text{is completely positive}\, \right\}\,.
\end{align}
Here \(\cJ_{V U^*}:\cB(\C^{n'}) \otimes \M_B \to \cB(\C^d) \otimes \cB(\cH) \) is the completely positive mapping
\begin{align}
\cJ_{V U^*}(x):=\tr_n\left[V U^* x \otimes \idty_{d'} U V^*\right] \,,
\end{align}
and \(\bar{\tr}_{n'}:\cB(\C^{n'}) \otimes \M_B \to \cB(\C^d) \otimes \cB(\cH)\) denotes the partial trace with respect to \(\C^{n'}\) together with tensoring the identity on \(\C^d\),
\begin{align}
\bar{\tr}_{n'}(x):= \sum_{i=1}^{n'} x_{ii} \otimes \idty_d\quad\mathrm{for}\quad(x)_{ij} \in \cB(\C^{n'}) \otimes \M_B\,.
\end{align}
\end{lemma}

We remark that \(c(V U^*)\) does not depend on the choice of the particular minimal Stinespring dilations $U,V$, as all of these are connected by either a unitary on \(\C^{d'}\) or on \(\C^d\). Thus, they either do not influence the mapping \(\cJ_{V^*U}\) or the mapping \(\bar{\tr}_{n'}\) and hence have no effect on the constant \( c(V U^*)\).

\begin{proof}[Proof of Lemma~\ref{thm:mainuncert}]
The proof relies on the ideas developed for finite-dimensional quantum systems~\cite{Tomamichel11}, and can be regarded as the dual version of it. Let $\H$ be a Hilbert space such that $\M_{ABC} \subset \cB(\H)$ is faithfully embedded and there exists a purifying vector $\ket{\psi} \in \H$ for $\omega_{ABC}$, that is, $\omega_{ABC}(x) = \Scp{\psi}{x \, \psi}$. We denote by $U:\H \to \C^{n'} \otimes \C^{d'} \otimes \H$ and $V: \H \to \C^{n} \otimes \C^{d} \otimes \H$ the isometries of the minimal Stinespring dilations corresponding to $\cE$ and $\cG$, respectively (as explained in the discussion preceding the proposition). Since $\M_A \subset \M_C^\prime$, we have $\cG(\cB(\C^n)) \subset \M_C^\prime$, and by Arveson's commutant lifting theorem~\cite[Theorem 1.3.1]{Arveson_1969}, there exists a representation 
\begin{align}
\pi_C:\M_C \to \cB(\C^{n} \otimes \idty_d \otimes \idty_\cH)^\prime = \idty_n \otimes \cB(\C^{d}) \otimes \cB(\H)  
\end{align}
such that we have $\pi_C(y)V = V y$ for $y \in \M_C$. It follows that the map $\tilde{\cG}:\cB(\C^n) \otimes \M_C \to \cB(\H)$ defined by 
\begin{align}
\tilde{\cG}(x \otimes y) = V^*(x \otimes \idty_d \otimes \pi_C(y))V = \cG(x)y
\end{align}
for $x \in \cB(\C^n)$, $y \in \M_C$ extends to a completely positive unital map $\tilde{\cG}:\cB(\C^n) \otimes \M_C \to \M_{AC}$. Due to the fact that $\ket{\psi}$ is a purification of $\omega_{ABC}$, we have that
\begin{align}
\Scp{V\psi}{x \otimes \idty_d V\psi} = \Scp{\psi}{\tilde{\cG}(x)\psi} = \omega_{ABC}(\tilde{\cG}(x)) = \omega_{GC}(x)
\end{align}
for $x \in \M_{GC} \cong \cB(\C^{n}) \otimes \M_B$, implying that $V\ket{\psi}$ is a purification of $\omega_{GC}$ on $\C^{n} \otimes \C^d \otimes \H$ with representation given by $\textrm{id}_{\cB(\C^{n})} \otimes \pi_C$, with $\textrm{id}_{\cB(\C^{n})}$ being the defining representation of $\cB(\C^{n})$ on $\mathbb{C}^{n}$. Since the commutant of $\cB(\C^n)$ in $\cB(\C^{n} \otimes \C^{d} \otimes \H)$ equals $\idty_n \otimes \cB(\C^{d} \otimes \H)$, the complementary system is computed as the commutant $\M_D=\pi_C(\M_C)^\prime \cap \cB(\C^d \otimes \H)$ of $\pi_C(\M_C)$ in $\cB(\C^d \otimes \H)$. An analogues argument constructs a channel $\tilde{\cE}:\cB(\C^{n'}) \otimes \M_B \to \M_{AB}$ starting from $\cE$, providing a purification $U\ket{\psi}$ of $\omega_{EB}$ on  $\C^{n'} \otimes \C^{d'} \otimes \H$ with complementary system $\M_{\tilde{D}}$. Here we denoted $\M_{\tilde{D}} = \pi_B(\M_B)^\prime$, with $\pi_B$ being the representation of $\M_B$ obtained from repeating the above arguments for $\cE$. Since by definition of the smooth conditional max-entropy (Definition~\ref{def:SmoothHmax}) 
\begin{align}
\SHmax{G}{C}_{\omega} = - \SHmin{G}{D}_{V\ket{\psi}} \, ,
\end{align}
we have to show that
\begin{align}
\SHmin{G}{D}_{V\ket{\psi}} \leq \SHmin{E}{B}_{U \ket{\psi}} + \log c\quad\text{where $c= c(V^* U)$ as in~\eqref{eq:choi}}\,.
\end{align}

We first prove the proposition for $\epsilon = 0$. By the operational characterization of the conditional min-entropy (Corollary~\ref{prop:HminDualForm}) the last inequality amounts to
\begin{align}
&\sup\Big\{\Scp{\psi}{U^* x \otimes \idty_{d'} \, U \psi} \,:\, x \in (\cB(\C^{n'}) \otimes \M_B)_+\,,\tr_{n'}(x) \leq \idty_\H \Big\}\notag\\
&\leq c\cdot\sup\Big\{ \Scp{\psi}{V^* \,y\, V \psi} \,:\, y \in (\cB(\C^{n}) \otimes \M_{D})_+\,,\tr_{n}(y) \leq \idty_{d} \otimes \idty_\H \Big\}\,.
\end{align}
Since $V^*V$ projects onto $\H$ and $U\ket{\psi} = U V^* V \ket{\psi}$, we have
\begin{align}
  \Scp{\psi}{U^* x \otimes \idty_{d'} \, U \psi} = \Scp{\psi}{V^*VU^* x \otimes \idty_{d'} \, UV^*V \psi}\,.
\end{align}
Let us now consider the expression $VU^* x \otimes \idty_{d'} \, UV^*$. If this would be an element of $(\cB(\C^{n}) \otimes \M_{D})_+$, the assertion would follow from
\begin{align}\label{eq:uncertmainineq}
  \tr_{n}(VU^*\, x \otimes \idty_{d'} \, UV^*) \;\leq\; c \cdot \idty_{d} \otimes \tr_{n'}(x)\,,
\end{align}
where $x \in \cB(\C^{n'}) \otimes \M_B$. However, this follows directly from the definition of the constant \(c(V U^*)\), so only the assumption needs to be checked. For that, note that since $\tilde{\cE}$ maps into $\M_{AB} \subset \M_C^\prime$, again by Arveson's commutant lifting theorem we can find a representation $\tilde{\pi}_C : \M_C \to (\textrm{id}_{\cB(\C^{n'})} \otimes \pi_B(\cB(\C^{n'}) \otimes \M_B))^\prime \subset \cB(\C^{n'd'}\otimes\H)$ satisfying $Uy=\tilde{\pi}_C(y)U$ and hence
\begin{align}
  VU^* x \otimes \idty_{d'} \, UV^* \pi_C(y)= VU^* x \otimes \idty_{d'} \, U y V^*&= VU^*  \tilde{\pi}_C(y) x \otimes \idty_{d'} \,U V^*\\
  &= \pi_C(y) VU^* x \otimes \idty_{d'} \, UV^*\,,
\end{align}
which implies $VU^* x \otimes \idty_{d'} \, UV^* \in \pi_C(\M_C)^\prime = \cB(\C^{n}) \otimes \M_{D}$. Since $x \in (\cB(\C^{n'}) \otimes \M_B)_+$, the expression $VU^* x \otimes \idty_{d'} \, UV^*$ also defines a positive operator. This concludes the proof for $\epsilon=0$.

For $\epsilon > 0$, take $\gamma^\epsilon_{GD}\in\Ball{}{\gamma_{GD}}$, where $\gamma_{GD}$ denotes the vector state $V \ket{\psi}$ restricted to $\cB(\C^{n})\otimes\M_{D}$. Let $(\pi_{GD},\cK)$ be a representation $\pi_{GD}:\cB(\C^{n})\otimes\M_{D} \to \cB(\cK)$ on $\cK$ such that there exists purifying vectors $\ket{\xi}$ and $\ket{\xi^\eps}$ for $\gamma_{GD}$ and $\gamma^\eps_{GD}$, respectively, with $\cF(\ket{\xi},\ket{\xi^\epsilon}) \geq 1 - \epsilon^2$. Moreover, there exists an isometry $W:\C^{nd} \otimes \H \to \cK$ satisfying $W x = \pi_{GD}(x)W$ for $x \in \cB(\C^{n})\otimes\M_{D}$ and $WV\ket{\psi} = \ket{\xi}$. We find using that the purified distance is monotone under partial isometries 
\begin{align}
  \cF(U \ket{\psi}, U V^* W^*\ket{\xi^\epsilon}) = \cF(U V^*V\ket{\psi}, U V^* W^*\ket{\xi^\epsilon}) &= \cF(U V^*W^*\ket{\xi}, U V^* W^*\ket{\xi^\epsilon})\\
  &\geq \cF(W^*\ket{\xi}, W^*\ket{\xi^\epsilon})\\
  &\geq 1-\eps^2\,,
\end{align}
and hence $U V^* W^*\ket{\xi^\epsilon}_{EB} \in \Ball{}{U \ket{\psi}_{EB}}$, where $U V^* W^*\ket{\xi^\epsilon}_{EB}$ (resp. $U \ket{\psi}_{EB}$) denotes the state on $\M_{EB}$ induced by the vector $U V^* W^*\ket{\xi^\epsilon}$ (resp. $U \ket{\psi}$). Moreover, we find for any $y= VU^* x \otimes \idty_{d'} \, UV^*$ with $x \in (\cB(\C^{n'}) \otimes \M_B)_+$ that 
\begin{align}
    \label{eq:helperrelationuncert}
    \Scp{W^*\xi^\eps} {y W^*\xi^\eps} 
    = \Scp{\xi^\eps}{\sqrt{\pi_{GD}( y)}WW^*\sqrt{\pi_{GD}(y)} \xi^\eps}
  \leq \Scp{\xi^\eps}{\pi_{GD}(y)\xi^\eps}
  = \gamma_{GD}^\eps(y) \,,
\end{align}
since $y=VU^* x \otimes \idty_{d'} \, UV^* \in (\cB(\C^{n}) \otimes \M_{D})_+$ as before. Thus, repeating the steps for the $\eps=0$ case and using~\eqref{eq:helperrelationuncert} yields the assertion.
\end{proof}

Finally, we obtain the proof of Proposition~\ref{thm:uncert} from Lemma~\ref{thm:mainuncert}.

\begin{proof}[Proof of Proposition~\ref{thm:uncert}]
  Since a measurement is a quantum channel with domain being an abelian von Neumann algebra we can make use of Lemma~\ref{thm:mainuncert}. Assume for simplicity that $|X|=|Y|=n$, and think of $\ell_n^\infty$ as the subalgebra of diagonal matrices in $\cB(\C^n)$. We then define the maps $\cG:\cB(\C^n) \to \M_A$, $\cE:\cB(\C^{n}) \to \M_A$ as being the projection onto the subalgebra of diagonal matrices followed by the measurement,
  \begin{align}
\cG\left(\sum_{x,x'} a_{x,x'} \ketbra{x}{x'}\right) = a_{x,x} E_A^x\,,
  \end{align}
  for $\sum_{x,x'} a_{x,x'} \ketbra{x}{x'} \in \cB(\C^n)$ and correspondingly for $\cE$. A corresponding isometry for $E:\ell^{\infty}_{n}\to\M_A$, $e_x \mapsto E_A^x$ can then be chosen of the form
\begin{align}
V \,:\,\H \to \C^n \otimes \C^n \otimes \H\quad\mathrm{with}\quad V\ket{\psi}:=\sum_{i=x}^n (E_A^x)^{\frac{1}{2}}\ket{\psi} \ket{x}\ket{x}\,,
\end{align}
and analogously \(U\) for $F:\ell^{\infty}_{n}\to\M_A$, $e_y \mapsto F_A^y$. Here $\{e_x\}$ denotes the canonical basis for $\ell^\infty_n$. However, these isometries are generally not minimal. This problem can be resolved by projecting onto the span of the respective representations. Let $P$ ($Q$) be the projector onto the subspace of $\C^n \otimes \C^n \otimes \H$ spanned by $aV\ket{\psi}$ ($aU\ket{\psi}$), for $a\in \cB(\C^n)$ and $\ket{\psi} \in \H$. It then follows that $PV$ and $QU$ are minimal Stinespring dilations.

For $A \in (\cB(\C^n) \otimes \M_B)_+$, we find that
\begin{align}
  \tr_n\left[PV U^*Q (A \otimes \idty_{d'}) QU V^*P\right] = \tr_n\left[PV \tilde{\cE}(A \otimes \idty_{d'})V^*P\right] = \tr_n\left[PV \sum_{y\in Y} F_A^y \,A_{yy} V^*P\right]\,,
\end{align}
where we used the extension $\tilde{\cE}:\cB(\C^n) \otimes \M_B \to \M_{AB}$ of $\cE$ constructed in the proof of Lemma~\ref{thm:mainuncert}. Moreover, for any \(\sigma \in \cN^+(\cB(\C^n\otimes \cH))\), \(\sigma = (\sigma)_{xx'}\), we have
\begin{align}
  (\sigma \otimes \tr_n)\left(PV \sum_{y\in Y} F_A^y \,A_{yy} V^*P\right) = (\sigma \otimes \tr_n)\left(PV \sum_{y\in Y} F_A^y \,A_{yy} V^*P\right)\,.
\end{align}
Since $\cG(\cB(\C^n)) \subset \M_A \subset \M_B^\prime$, we can find by Arveson's commutant lifting theorem~\cite[Theorem 1.3.1]{Arveson_1969} a representation $\pi_B :\M_B \to (\cB(\C^n) \otimes \idty_n \otimes \idty_\H)^\prime = \idty_n \otimes \cB(\C^n) \otimes \cB(\H)$ such that $PVb=P\pi_B(b)V=\pi_B(b)PV$. This shows that $P$ both commutes with $V \sum_{y\in Y} F_A^y \, V^*$ (by construction) as well as with $\pi_B(\M_B)$ and we find
\begin{align}
  (\sigma \otimes \tr_n)\left(PV \sum_{y\in Y} F_A^y \,A_{yy} V^*P\right) &= (\sigma \otimes \tr_n)\left(PV \sum_{y\in Y} F_A^y \, V^*P \pi_B(A_{yy})\right)\\
  &\leq (\sigma \otimes \tr_n)\left(V \sum_{y\in Y} F_A^y \,A_{yy} V^*\right)\\
  &= \sum_{x,y=1}^n \sigma_{xx}((E_{A}^{x})^{\frac{1}{2}} F_{A}^{y} (E_{A}^{x})^{\frac{1}{2}} A_{yy})\\
  &\leq\max_{x,y} \left\|(E_{A}^{x})^{\frac{1}{2}} F_{A}^{y} (E_{A}^{x})^{\frac{1}{2}}\right\|\cdot\sum_{x,y=1}^n \sigma_{xx}( A_{yy})\label{eq:helperuncertmeas}\,.
\end{align}
The result follows since~\eqref{eq:helperuncertmeas} implies the bound
\begin{align}
c(V U^*) \leq \max_{x,y} \left\|(E_{A}^{x})^{\frac{1}{2}}\cdot(F_{A}^{y})^{\frac{1}{2}}\right\|^{2}\,.
\end{align}
\end{proof}


\subsection{Quantum Key Distribution}\label{sec:qkd_applications}

One goal in quantum information theory is a tight characterization of information-theoretic tasks involving quantum systems. We focus on the particular task of quantum key distribution for which the basic information-theoretic tasks can be characterized by smooth conditional min- and max-entropies if only finite-dimensional spaces are involved~\cite{renner05}. We prove that this remains true even when quantum systems are modeled by von Neumann algebras. We first describe the task of quantum key distribution and divide it into two subtasks, which are then characterized by the smooth conditional min- and max-entropy.

We consider a tripartite setting with space-like separated parties Alice $(A)$, Bob $(B)$, and Eve $(E)$. The goal for Alice and Bob is to create a uniformly distributed random bit string, the key, which is known to both of them (correctness condition), but not to the adversary Eve (security condition). Mathematically, we model Alice and Bob as a bipartite system $\M_{AB}=\M_{A}\vee\M_{B}$ with von Neumann algebras $\M_{A}$ and $\M_{B}$, and denote the state they share by $\omega_{AB}\in\cS(\cM_{AB})$. Furthermore, we assign to Eve the complementary system $\cM_E$ of a purification $\omega_{ABE}$ of $\omega_{AB}$. After Alice measured her system by applying some POVM $\{E_{A}^{x}\}_{x\in X}\subset\cM_{A}$, $|X|<\infty$ the resulting post-measurement state is modeled by a classical quantum state $\omega_{XBE} \in \S(\ell^{\infty}_X\otimes\M_{BE})$. Bob then wants to determine Alice's bit string and for that he receives a classical message $M$ from Alice. Based on this, Bob chooses his measurement to optimize the success probability to obtain the same bit string. This task is known as data compression with quantum side information and was linked to the smooth conditional max-entropy in finite dimensions~\cite[Theorem 1]{renes10}. In the last step Alice and Bob extract a secure key from the bit string they share. This is referred to as privacy amplification and in finite dimensions it has been shown that the remaining correlation with Eve's system after this step can be quantified by the smooth conditional min-entropy~\cite[Theorem 6]{tomamichel10}. In the following, we discuss these two information-theoretic tasks in detail for our more generalized setting.\\

\textbf{Privacy amplification against quantum side information.} We commence with a classical quantum state $\omega_{XE}\in\cS(\ell^\infty_X \otimes \M_E)$ between Alice and Eve. As outlined in the introduction, the task of privacy amplification is to extract a secure key from $\omega_{XE}$, that is, a uniformly distributed bit string $K$ on Alice's side that is uncorrelated with Eve's system $E$. This is described by a classical quantum state $\frac{1}{|K|}\tau_{K}\otimes\sigma_{E}$, where $\frac{1}{|K|}\tau_{K}$ is the tracial state on $\ell_K^{\infty}$ and $\sigma_{E}\in\cS(\cM_{E})$. Note that $K$ is a classical random variable generated from $X$. We follow~\cite{renner05} and call a state $\omega_{KE}\in \cS_{\leq}(\ell^{\infty}_K\otimes\cM_{E})$ $\epsilon$-secure if
\begin{align}
\left\|\omega_{KE}-\frac{1}{|K|}\tau_{K}\otimes\omega_{E}\right\|\leq\epsilon\,.
\end{align}

The basic idea how to achieve an $\epsilon$-secure key from an input $\omega_{XE}$ is to randomly combine several indices $x$ into a single one, and thereby reducing (hashing) the alphabet from $X$ to $K$ with $|K|<|X|$. This process can be accomplished by using two-universal hash functions. A family of $\{X,K\}$-hash functions is a set $\{ \F,\P_\F \}_{X,K}$, where every element $f \in \F$ is a function $f: X \to K$, called hash function, and $\P_\F$ is a probability measure on the set $\F$. A family of $\{X,K\}$-hash functions is called two-universal if for all $x,y \in X$ with $x \neq y$
\begin{align}\label{def,eq:twouni}
\P_\F \left( f(x) = f(y) \right) \leq \frac{1}{|K|}\,.
\end{align}
We refer to~\cite{Carter79,Wegman81} for the existence proof of families of two-universal $\{X,K\}$-hash functions for every two finite alphabets $X,K$ with $|K|\leq|X|$. Given a hash function $f:X\rightarrow K$, we define the operator $T_f$ from $\cS_{\leq}(\ell^{\infty}_X)$ to $\cS_{\leq}(\ell^{\infty}_K)$ through
\begin{align}\label{eq:hashops}
(T_fu)(i):=\sum_{x \in X: f(x)=i} u(x)\quad\mathrm{for}\quad u \in\cS_{\leq}(\ell^{\infty}_X)\quad\mathrm{and}\quad i \in K\,,
\end{align}
which implements the action of the hash function on the state space. We are now ready to state the main result of this section.

\begin{proposition}\label{thm:privampl}
Let $X,K$ be sets of finite cardinality with $|K|\leq|X|$, $\{ \F,\P_\F \}_{X,K}$ a family of two-universal $\{X,K\}$-hash functions, $\omega_{XE}=(\omega_{E}^{x})_{x\in X}\in \cS_{\leq}(\ell^{\infty}_X\otimes\cM_{E})$, and $\epsilon \geq 0$. Denoting by $\E_\F$ the expectation with respect to $\P_\F$, we have
\begin{align}\label{thm,eq:privampl}
\E_\F \left\|(T_f \otimes \mathcal{I})( \omega_{XE}) - \frac{1}{|K|} \tau_{K} \otimes \omega_E\right\| \leq \sqrt{|K|\cdot2^{-\mathrm{H}_{\min}^{\epsilon}(X|E)_{\omega}}}+4\epsilon\,.
\end{align}
\end{proposition}

We note that our proof is different from the one for finite-dimensional systems presented in~\cite{tomamichel10}. Our proof strategy is inspired by the purely classical results~\cite{Bennett88,Levin89,Bennett95}. We show the statement for $\epsilon = 0$, from which the $\epsilon > 0$ case is obtained by a simple application of the triangle inequality (see~\cite[Section 5.6]{renner05} for details).

\begin{proof}
Recall that the norm on $\ell^1(\cN(\cM_{E}))$ is inherited from the dual of $\ell^{\infty}_X \otimes \M_E$, such that the left hand side of~\eqref{thm,eq:privampl} is simply the expectation value $\E_\F$ of
\begin{align}\label{eq:privampnorm}
\sum_{i \in K} \sup_{\substack{a_i \in \M_E}{\norm{a_i}=1}} \left| \sum_{x \in X: f(x)=i} \omega_E^x(a_i) - \frac{1}{|K|}\omega_E(a_i) \right|\,.
\end{align}
Because $\vert X\vert$ is finite, we can assume that there exists a $\sigma_{E}\in\cS(\M_E)$ such that $\omega_E^x \leq \lambda\cdot\sigma_{E}$ for all $x \in X$ and suitable $\lambda >0$ (take for instance $\sigma_E=\sum_x \omega_E^x$). We choose $(\pi_\sigma,\H_\sigma, \ket{\xi_\sigma})$ to be a purification of $\sigma_E$ such that $\ket{\xi_\sigma}$ is cyclic. This is always possible according to the GNS construction. We denote by $D_x\in \pi_\sigma(\M_E)'$ ($D\in \pi_\sigma(\M_E)'$) the corresponding Radon-Nikodym derivatives~\cite[Chapter VII.2]{Takesaki1} of $\omega^x_E$ ($\omega_E$) with respect to $\sigma_{E}$. That is, \(D_x \ket{\xi_\sigma}\) (\(D\ket{\xi_\sigma}\)) is a purification of \(\omega^x_E\) (\(\omega_E\)) and we have \(\omega_E^x \leq \norm{D_x}^2 \sigma\). Since $\ket{\xi_\sigma}$ is cyclic it follows that $\sum_x D_x^* D_x \ket{\xi_\sigma} = D^* D \ket{\xi_\sigma}$. We can then write
\begin{align}
\sum_{x \in X: f(x)=i} \omega_E^x(a_i) - \frac{1}{|K|}\omega_E(a_i) = \left\langle\left( \sum_{x \in X: f(x)=i} D_x^* D_x - \frac{1}{|K|} D^* D \right) \xi_\sigma\Big|\pi_\sigma(a_i) \xi_\sigma\right\rangle\,,
\end{align}
where we used the fact that $D_x$ as well as $D$ are elements of the commutant of $\pi_\sigma(\M_E)$. We now insert this expression into~\eqref{eq:privampnorm}, take the expectation $\E_\F$ and apply the Cauchy-Schwarz inequality to the sum over $i\in K$ and $f \in\cF$, which yields
\begin{align}
\E_\F \Big\|(T_f\otimes\mathcal{I})(\omega_{XE})&-\frac{1}{|K|} \tau_{K} \otimes \omega_E\Big\|\notag\\
&\leq \sqrt{|K|}\left( \E_\F \sum_{i \in K}\left\langle\xi_\sigma\Big|\left( \sum_{x \in X: f(x)=i} D_x^* D_x - \frac{1}{|K|} D^* D \right)^2\xi_\sigma\right\rangle\right)^{\frac{1}{2}}\,.
\end{align}
Using that $\E_\F \sum_{i \in K}\sum_{x \in X: f(x)=i}=\sum_{x \in X}$ and the identity $\sum_x D_x^* D_x \ket{\xi_\sigma} = D^* D \ket{\xi_\sigma}$, we can compute
\begin{align}\label{eq:privampbinomial}
& \E_\F \sum_{i \in K}\left\langle\xi_\sigma\Big|\left( \sum_{x \in X: f(x)=i} D_x^* D_x - \frac{1}{|K|} D^* D \right)^2\xi_\sigma\right\rangle \nonumber\\
&= \E_\F \sum_{i \in K}\left\langle\xi_\sigma\Big|\left( \sum_{x \in X: f(x)=i} D_x^* D_x \right)^2 \xi_\sigma\right\rangle - \frac{1}{|K|}\left\langle\xi_\sigma|D^* D D^* D \xi_\sigma\right\rangle\,.
\end{align}
The sum in the first term can be written as
\begin{align}
\E_\F\sum_{i \in K}\Big( \sum_{x \in X: f(x)=i} D_x^* D_x \Big)^2 = &\sum_{x \in X} \sum_{y \in X}D_x^* D_x W_{xy}D_y^* D_y\,,
\end{align}
where $W_{xy}:=\E_\F\sum_{i \in K}\, \delta_{f(x)=i}\; \delta_{f(y)=i}$. Note that this defines a positive $|X| \times |X|$-matrix which can be upper bounded by $P_{\tau_X} + \idty_X$, with $\idty_X$ the $|X| \times |X|$-identity matrix and $P_{\tau_X}$ the projector onto the vector corresponding to the uniform distribution on $X$, normalized to trace one. This follows from the definition of two-universal hash functions~\eqref{def,eq:twouni}. Using these facts, we obtain
\begin{align}
\E_\F \sum_{i \in K}\left\langle\xi_\sigma\Big|\left( \sum_{x \in X: f(x)=i} D_x^* D_x - \frac{1}{|K|} D^* D \right)^2\xi_\sigma\right\rangle \leq \sum_{x \in X} \Scp{\xi_\sigma}{D_x^* D_x D_x^* D_x \xi_\sigma}\,.
\end{align}
The expression on the right hand side can be estimated further by employing $\Scp{\xi_\sigma}{\sum_{x \in X} D_x^* D_x \xi_\sigma} = \omega_E(\idty) \leq 1$. Hence, we find
\begin{align}
\sum_{x \in X} \Scp{\xi_\sigma}{D_x^* D_x D_x^* D_x \xi_\sigma}\leq\max_{x\in X}\;\left\|D_x^*D_x\right\|\;\left\langle\xi_\sigma\Big|\sum_{x\in X}D_x^*D_x\xi_\sigma\right\rangle\leq 2^{-\Hmin{X}{E}_{\omega}}\,,
\end{align}
where we also took the infimum over all suitable $\sigma_{E}\in\cS(\M_{E})$ that majorize every $\omega^x_{E}$. Putting the steps together, we arrive at
\begin{align}
\E_\F\left\|(T_f\otimes\mathcal{I})(\omega_{XE})-\frac{1}{|K|}\tau_{K}\otimes\omega_E\right\|\leq\sqrt{|K|\cdot 2^{-\Hmin{X}{E}_{\omega}}}\,.
\end{align}
\end{proof}

\textbf{Data compression with quantum side information.} We consider a classical random variable $X$ correlated to a quantum state on a von Neumann algebra $\cM_{B}$. This is modeled by a classical quantum state $\omega_{XB} \in \S(\ell^{\infty}_{X}\otimes\M_B)$. A one-way classical communication protocol to transmit $X$ from Alice to Bob consists of a classical encoding map $\EE:\ell^{\infty}_C \rightarrow \ell^{\infty}_X$ on Alice's side, and a decoding map $\DD:\ell^{\infty}_X \rightarrow \ell^{\infty}_C\otimes\M_B$ on Bob's side, where $\EE$ and $\DD$ are quantum channels. The classical alphabet $C$ (code space) specifies the number of bits, $\log|C|$, that are transmitted. The pre-dual of the decoding map can be written as $\DD_* = \{\DD_*^c\}_{c\in C}$, where the map $\DD_*^c$ onto the classical outcome $X$ is described by a POVM $ \{D^c_x\}_{x\in X}$. In the following every such protocol is specified by the triple $(\EE,\DD,C)$.
\begin{definition}
Let $X$ be a set of finite cardinality and $\omega_{XB}\in\S(\ell^{\infty}_{X}\otimes\M_B)$. The error probability of a protocol $(\EE,\DD,C)$ for $\omega_{XB}$ is defined as
\begin{align}
p_{\mathrm{err}}(\omega_{XB};\EE,\DD):= 1 - \sum_{x} \omega^x_B(D_x^{\EE_* (x)})\,,
\end{align}
where $D_x^{\EE_* (x)}= \sum_c ( \EE_*(x))_c  D^c_x$.
\end{definition}

The main result is the following quantification of the achievable error probability.

\begin{proposition}\label{cor:SmoothDataCompr}
Let $X$ be a set of cardinality $|X|$, $\omega_{XB}\in\S(\ell^{\infty}_{X}\otimes\M_B)$, and $\epsilon\geq0$. Then, there exist for any alphabet $C$ with $|C|\leq|X|$ an encoding map $\EE$ and a decoding map $\DD$, such that the protocol $(\EE,\DD,C)$ satisfies
\begin{align}\label{cor,eq:SmoothDataCompr}
p_{\mathrm{err}}(\omega_{XB};\EE,\DD) \leq \sqrt{\frac{1}{|C|}\cdot 2^{\SHmax{X}{B}_{\omega} + 3}} + 2\epsilon\,.
\end{align}
\end{proposition}

Our proof is along the line of the arguments for quantum side information modeled by finite-dimensional spaces~\cite[Theorem 1]{renes10}. In particular, for the encoding we employ the property of a family of two-universal hash functions $\F$ as in~\eqref{def,eq:twouni}. We show that the averaged error probability over a family of two-universal hash functions $\F$ is bounded as in~\eqref{cor,eq:SmoothDataCompr}, and from this we can then conclude that there exists a function $f\in\F$ suitable as an encoding map. Now assume that Alice holds the value $x$ and sends the message $c=f(x)$ to Bob. Bob then knows that $x\in f^{-1}(c)$, and applies as the decoding map a measurement which is appropriate to distinguish between the states $\omega_B^x$ for $x\in f^{-1}(c)$. For that, he uses a POVM $\{ D_{x';f}^c\}_{x'\in X}$ with $D_{x';f}^c=0$ if $x'\notin f^{-1}(c)$, which we choose as an adapted pretty good measurement to distinguish the ensemble $\{\omega_B^x\}_{x\in f^{-1}(c)}$~\cite{PrettyGoodMeasurement}. 
Adapted pretty good measurement means that we have to add $\epsilon \idty$ ($\epsilon >0$) to certain operators in order to take their inverse. Eventually, we take the limit $\epsilon\rightarrow 0$. 

The error analysis in the finite-dimensional case is crucially based on an operator inequality from~\cite{hayashi03,Auedenaert2007}, whereas we use the following generalization to von Neumann algebras.
\begin{lemma}\cite[Proposition 1.1]{Ogata2010}
Let $\phi,\eta \in \Ss(\M)$, $s_+$ be the support projection onto the positive part of $\phi-\eta$, and $s_-=\idty-s_+$. Then, we have
\begin{align}
\phi(s_-) + \eta(s_+) \leq \cF_{\cM}(\phi,\eta)^{\frac{1}{2}}\,.
\end{align}
\end{lemma}


\section{Discussion and Outlook}\label{sec:outlook}

We generalized the smooth entropy formalism to von Neumann algebras and discussed various properties in this framework. We showed that the characterizations of privacy amplification and data compression in terms of the smooth conditional min- and max-entropy still hold. The results in this paper can be used to extend one-shot quantum information-theoretic tasks to more general quantum systems described by continuous variables and in particular fermionic and bosonic quantum fields. For example, by building on the results given here, we prove security of a squeezed state continuous variable quantum key distribution protocol~\cite{furrer12,furrer13}. 
Since the smooth min- and max-entropy have also been used in thermodynamics (see, e.g., \cite{delrio11}), the generalization to the von Neumann algebra setting is also interesting from a physical perspective. Especially as quantum mechanical systems of interest in thermodynamics often possess an infinite number of degrees of freedom. One could also generalize the formalism for quantum side information to operator systems~\cite{PaulsenOpSys}. Operationally, this corresponds to a restriction of the actual measurements that are allowed to perform on the physical system. This restriction could be conducted at a fundamental level, by excluding the elements of the von Neumann algebra that are unphysical in the sense that they cannot be observed. For a task like data compression with quantum side information, this would allow to constraint the quantum measurements at the decoder.


\section*{Acknowledgments}

We thank Renato Renner for instructive discussions about privacy amplification. We would also like to thank Marco Tomamichel for many insightful discussions about the smooth entropy formalism, and for detailed feedback on the first version of this paper. We acknowledge discussions with Matthias Christandl, Reinhard F.~Werner, Michael Walter, and Joseph M.~Renes. We thank an anonymous reviewer for pointing out an error in the proof of Lemma~\ref{thm:mainuncert} and a detailed explanation of how to fix it. MB and VBS are both grateful for the hospitality and the inspiring working environment at the Institute Mittag-Leffler in Djursholm, Sweden, where this work was started. Most of this work was done while MB was at ETH Zurich, and FF and VBS were at the University of Hanover. MB acknowledges funding provided by the Institute for Quantum Information and Matter, an NSF Physics Frontiers Center (NFS Grant PHY-1125565) with support of the Gordon and Betty Moore Foundation (GBMF-12500028). Additional funding support was provided by the ARO grant for Research on Quantum Algorithms at the IQIM (W911NF-12-1-0521). FF acknowledges support from the Graduiertenkolleg 1463 of the Leibniz University Hanover and by the Japan Society for the Promotion of Science (JSPS) by KAKENHI grant No. 24-02793, and FF and VBS both acknowledge support by the BMBF project QUOREP as well as the DFG cluster of excellence QUEST.


%

\end{document}